\documentclass[aps, pra, twocolumn, notitlepage,superscriptaddress,nofootinbib   
]{revtex4-1}
\usepackage{amssymb,latexsym,amsmath}
\usepackage{bm}
\usepackage{amssymb,amsmath}
\usepackage{graphicx,xcolor}

\usepackage[caption=false,lofdepth,lotdepth]{subfig}

\usepackage{hyperref}   
\hypersetup{%
	pdfpagemode=None, 
	pdfstartpage=1,
	pdfmenubar=true,
	pdftoolbar=true,
	colorlinks = true,
	linkcolor=blue,
	citecolor=blue,
	urlcolor=blue,
	bookmarksopen=false
}

\newcommand{\beq}{\begin{equation}}   
\newcommand{\eeq}{\end{equation}}
\newcommand{\nn}{\nonumber}

\newcommand{\bs}{{\bm s}}
\newcommand{\bw}{{\mathbf{w}}}
\newcommand{\bn}{{\bm n}}

\newcommand{\jt}{\vartheta}
\newcommand{\mO}{\mathcal{O}}

\begin{document}
	
\title{Superfluid vortex dynamics on a torus and other toroidal surfaces of revolution}
\author {Nils-Eric Guenther}
\email{nils.guenther@icfo.eu}
\affiliation{ICFO -- Institut de Ci\`encies Fot\`oniques, The Barcelona Institute of Science and Technology, 08860 Castelldefels (Barcelona), Spain}
\author {Pietro Massignan}
\email{pietro.massignan@upc.edu}
\affiliation{Departament de F\'isica, Universitat Polit\`ecnica de Catalunya, Campus Nord B4-B5, E-08034 Barcelona, Spain}
\affiliation{ICFO -- Institut de Ci\`encies Fot\`oniques, The Barcelona Institute of Science and Technology, 08860 Castelldefels (Barcelona), Spain}
\author {Alexander L.\ Fetter}
\email{fetter@stanford.edu}
\affiliation {Departments of Physics and Applied Physics, Stanford University, Stanford, CA 94305-4045, USA}
\date{\today}

\begin{abstract}

The superfluid flow velocity 
is proportional to the gradient of the phase of the superfluid order parameter, leading to the quantization of circulation  
around a vortex core. In this work, we study the dynamics of a superfluid film on the surface of a torus.
Such a compact surface   allows only configurations of vortices with zero net vorticity. We derive analytic expressions for the flow field, the total energy, and the time-dependent dynamics of the vortex cores. 
The local curvature of the torus and  the presence of non-contractable loops on this multiply connected surface alter both the superfluid flow and the 
vortex dynamics. Finally we consider
more general surfaces of revolution, called toroids. 

\end{abstract}

\maketitle
	
\section{Introduction}

The concept of point vortices in a two-dimensional incompressible and inviscid fluid (also called a {\it perfect fluid}) originated in the seminal works of Helmholtz \cite{Helmholtz1858}. For such a fluid, the number of  point vortices is conserved and, once the effect of the container boundary has been included, the time-dependent evolution of the entire fluid flow field can be reduced completely to the dynamics of such point vortices. Moreover, the vortices can be described in the framework of classical Hamiltonian mechanics \cite{Kirchhoff1876,Lin1941-1,Lin1941-2}, with the two 
coordinates of each vortex forming a pair of canonical Hamiltonian variables. These remarkable properties have generated considerable interest not only among
physicists but also among 
 mathematicians, with a wide literature available (see for example \cite{Hassan2007}).

The model of a perfect fluid, initially studied as a simplification of classical fluid dynamics, is almost exactly realized in quantum superfluid phases, and the point-vortex model has found wide application in the context of superfluid helium. 
The same formalism also  applies directly to dilute ultracold superfluid atomic Bose-Einstein condensates (BECs),

which have the important advantage of unprecedented control over many experimental parameters, such as interactions and confining potentials~\cite{PethickSmith2008book,PitaevskiiStringari2016book}. 
A superfluid state is characterized by a complex scalar order parameter (a condensate wave function) $\Psi = \sqrt{\rho} e^{i\Phi}$,  where $\rho$ is the condensate number density. Here the phase $\Phi$ determines the superfluid velocity
\begin{align}
{\bf v} = (\hbar/M)\bm \nabla\Phi,\label{Phase_Flow}
\end{align}
where $M$ is the mass of the superfluid particles
and $\hbar$ is the reduced Planck's constant.
Equation \eqref{Phase_Flow}  implies immediately that the flow field is irrotational, apart 
from singular points.

 Although dilute ultracold superfluid BECs are compressible, local changes in the density can be neglected in the Thomas-Fermi (TF) limit, appropriate in  many typical experiments.  In the incompressibile limit the density $\rho$ is constant, and the condition of current conservation for steady flow  $\bm \nabla \cdot (\rho {\bf v}) = 0$ reduces to the condition $   \bm \nabla \cdot {\bf v }= 0 $, namely that the velocity field is divergence-less. In this case, one can  define a scalar stream function $\chi$ such that  
\begin{figure}[t!]
	\includegraphics[width=0.6\columnwidth]{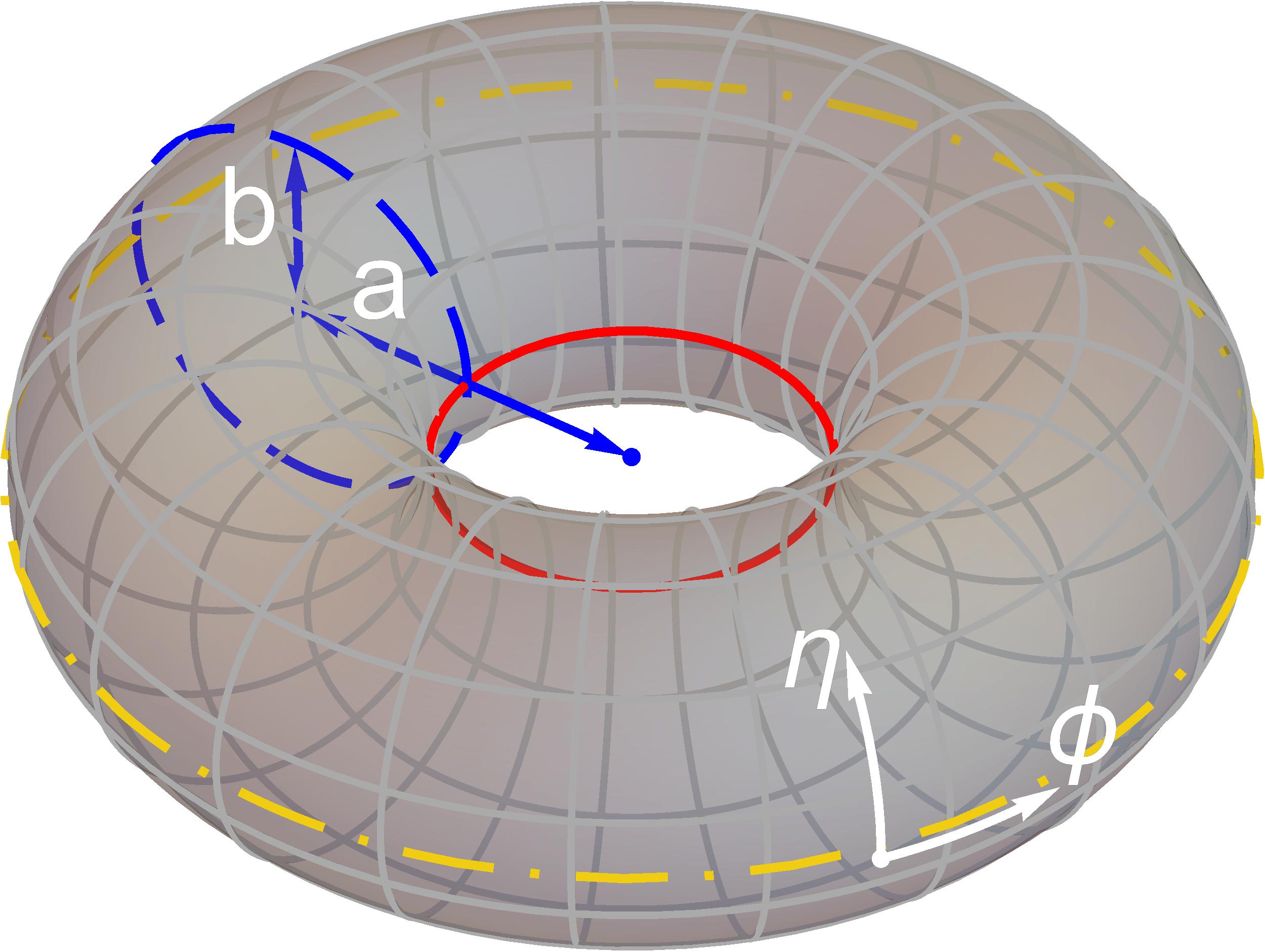} \\
	\vspace{10pt}
	\includegraphics[width=0.7\columnwidth]{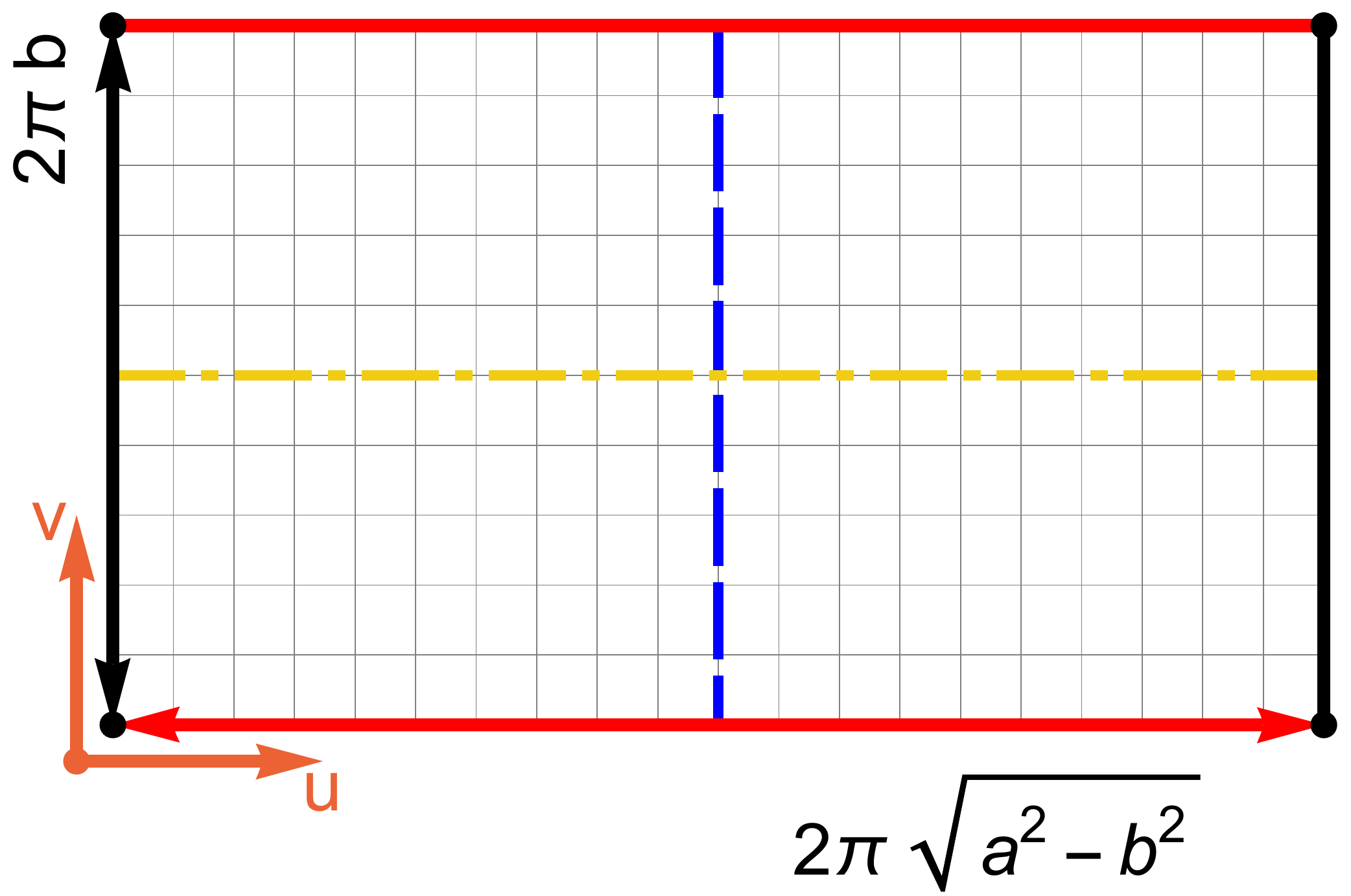}
	\caption{The surface of a torus with major radius $a$ and minor radius $b$ (top panel) can be conformally mapped to a doubly periodic cell with sides $2 \pi \sqrt{a^2-b^2}$ and $2\pi b$ (bottom panel). The red, blue and yellow orbits on the torus surface appear as straight lines in the $\{u,v\}$ cell.}
	\label{fig:ConformalMap}
\end{figure}
\begin{equation}\label{StreamFunction_Flow}
{\bf v} = (\hbar/M)\,\hat {\bm n}\times \bm \nabla \chi,
\end{equation}
\noindent with $\hat{\bm n} = \hat{\bm x}\times\hat{\bm y}$ the unit vector normal to the two-dimensional plane. Lines of constant $\chi$, called streamlines, are the trajectories of particles immersed in the fluid, giving the stream function a direct physical interpretation. The relations between $\Phi, \chi$ and the flow field 
\begin{equation}\label{CR}
\frac{M {\bf v}_x }{\hbar  } = \frac{\partial\Phi}{\partial x} = - \frac{\partial\chi}{\partial y},
\quad \frac{M {\bf v}_y}{\hbar} =  \frac{\partial\Phi}{\partial y} =  \frac{\partial\chi}{\partial x}
\end{equation}
can be interpreted as the Cauchy-Riemann equations for the real and imaginary parts of a complex function $F(z) = \chi + i\Phi$ with complex coordinate $z= x+ i y$. Together, these equations have the complex form
\begin{equation}\label{PotentialFlow}
{\bf v}_y + i {\bf v}_x = (\hbar/M)\,F'(z)
\end{equation} 
implying that $F(z)$ must be analytic almost everywhere. 

The description of two-dimensional flow in terms of a complex  potential has a long history  in classical hydrodynamics.  In a superfluid, however, the dynamics becomes quantized  because the velocity potential $\Phi$ has the physical significance of a condensate phase and must be uniquely defined, up to multiples of $2\pi$. The most famous consequence of this requirement is the quantization of vorticity, allowing only multiples of the elementary ``charge'' of vorticity $2 \pi \hbar/M$.

For a superfluid container with a nontrivial topology, this quantization condition leads to new constraints on the flow field, as  studied first by Ref.~\cite{Vinen1961} (see Ref.~\cite{Donnelly1991book} for a comprehensive summary). More recently, the vortex dynamics on the surfaces of cylinders and cones has been investigated in Refs.~\cite{Guenther2017,Massignan2019}.

A cylinder and a cone are both unbounded, similar to the infinite plane. Each of these configurations can support single vortices, because the lines of constant phase can escape to infinity. The situation is very different for a compact surface such as a sphere or a torus. Here the lines of constant phase emerging from a vortex must converge on another vortex with opposite sign, implying that the total vortex charge on a compact surface must vanish~\cite{Lamb1945book,Turner2010}. 
The high symmetry of the sphere means that the dynamics of a vortex dipole is particularly simple.
As noted briefly in Sec.~160 of  Lamb~\cite{Lamb1945book}, the problem can be solved straightforwardly via a stereographic projection.

In the present  work we study the dynamics of superfluid vortices on the surface of a torus, the simplest compact and multiply connected surface.  For such a curved two-dimensional surface, 
Eqs.~\eqref{Phase_Flow} and \eqref{StreamFunction_Flow} still hold, but now only in the local tangent space of the surface at any given point.
In Sec.~\ref{sec:MapTorus}, we introduce an {\it isothermal} set of coordinates, see Eq.~\eqref{LineElement_uv}, 
for the torus' surface, allowing an efficient description of the flow field induced by point vortices.
In Sec.~\ref{sec:FlowPotential} this flow is discussed in the framework of complex potentials. Section \ref{sec:Energy} derives the energy functional for vortex ensembles, and Sec.~\ref{sec:Dynamics} obtains the  time-dependent dynamics of the vortices themselves. Finally, Sec.~\ref{sec:GenralizedRS} extends the treatment in the previous sections to superfluid films on generalized toroidal surfaces, deriving an appropriate isothermal coordinate set.


\section{Isothermal coordinates}\label{sec:MapTorus}

A torus has a major radius $a$ and a minor radius $b$, with $a\geq b$, as shown in the top panel of Fig.~\ref{fig:ConformalMap}. Every point $\bs$ 
on  its surface can be parameterized in terms of two angles $\phi,\eta \in \{-\pi,\pi\}$ in Cartesian coordinates:
\begin{align}
\bs(\phi,\eta)=
\begin{cases}
x= \left(a+b\cos\eta\right)\cos \phi \\
y= \left(a+b\cos\eta\right)\sin \phi \\
z= b\sin \eta, 
\end{cases}
\label{TorusParametrization}
\end{align} 
where $\phi$ is the ``toroidal'' angle  and $\eta$ is the ``poloidal'' angle.
Correspondingly, the squared  line element $ds^2=dx^2+dy^2+dz^2$ on the surface becomes
\begin{align}
ds^2 = (a+b \cos\eta)^2 d\phi^2 + b^2 d\eta^2\equiv \lambda_\phi^2d\phi^2+\lambda_\eta^2d\eta^2. \label{LineElement_etaphi}
\end{align}  
The coordinate set $\{\phi,\eta\}$ is curvilinear, since physical distances are related to infinitesimal elements $d \phi$ and $d \eta$ through the local scale factors $\lambda_\phi$ and $\lambda_\eta$. This metric distorts the gradients appearing in Eqs.~\eqref{Phase_Flow} and \eqref{StreamFunction_Flow}, precluding a formulation in terms of a complex function $F(\phi + i \eta)$ of a complex variable $\phi + i \eta$.  

Such formulation is still possible, however, if one finds a suitable set of {\it isothermal} coordinates $\{u,v\}$ with squared line element 
\begin{align}
ds^2 = \lambda^2\left(du^2+dv^2\right), \label{LineElement_uv}
\end{align}

which ensures that the coordinates $\{u,v\}$ are a conformal parametrization of the surface. 
Kirchhoff already gave such an isothermal coordinate system for a torus 
 in 1875~\cite{Kirchhoff1875} in the context of
electrohydrodynamics on two-dimensional surfaces:
\begin{align}
	\phi = \frac{u}{c}, \quad \tan\left(\frac{\eta}{2}\right) = \sqrt{\frac{a+b}{a-b}} \tan\left(\frac{v}{2 b} \right)\label{KirchhoffTransformation},
\end{align}
with $c=\sqrt{a^2-b^2}$. Lengthy but elementary algebra shows  that $\{u,v\}$ are indeed isothermal and 
satisfy Eq.~\eqref{LineElement_uv} with local scale factor
\begin{align}
\lambda = \frac{c}{a-b \cos(v/b)}. \label{lambda_Torus}
\end{align}

In the $\{u,v\}$ plane, these coordinates vary in a rectangular cell of dimensions $\{-\pi c, \pi c \} \times \{-\pi b , \pi b \}$  with opposite edges identified.   The bottom panel of Fig.~\ref{fig:ConformalMap} illustrates this situation.
Section \ref{sec:GenralizedRS} derives the transformation given  in 
 Eq.~\eqref{KirchhoffTransformation},
 where we extend the analysis to more general  toroidal surfaces of revolution.


\begin{figure*}[t]
		\includegraphics[width=0.7\linewidth]{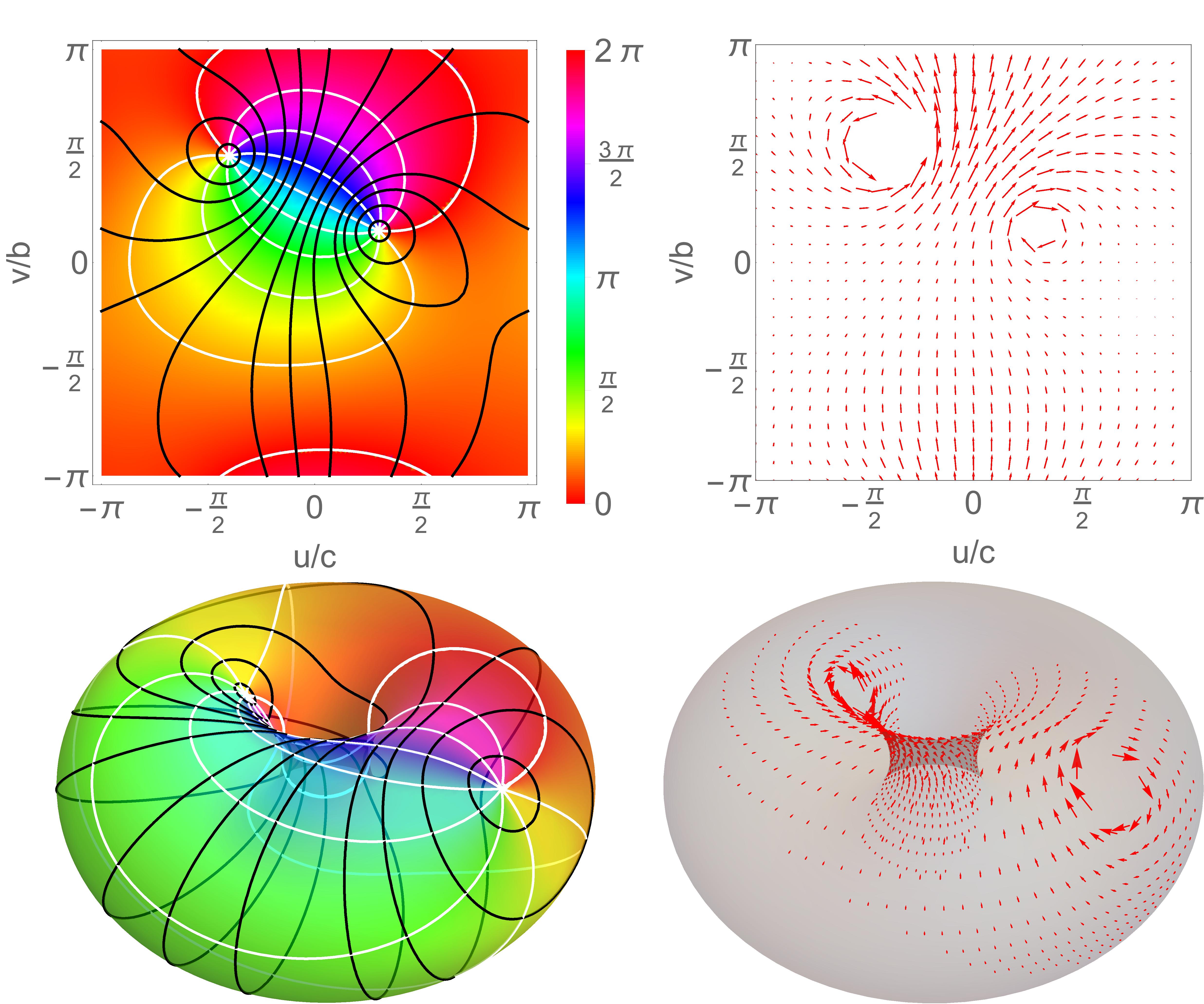}
			\caption{
				Phase (left panels) and corresponding flow field (right panels) for a torus with major radius $a$ and minor radius $b$ such that $a=\sqrt{2}b$, depicted in the $\{u,v\}$ coordinate set (top panels) and schematically mapped to the surface of the torus (bottom panels). The black lines in the left panels indicate stream lines, the white lines curves of constant phase. For clarity, the 
				right panels omit points too close to the vortex cores, as well as regions with vanishing velocity.
				\label{fig:FlowPotential_Dipole}}
\end{figure*}


\section{Complex flow potential of vortex configurations} \label{sec:FlowPotential}

With the isothermal coordinate set $\{u,v\}$ established in the last section, we can find the flow field in terms of a complex potential function. Define the complex coordinate $w=u+i v$ describing the point $\bs$ on the torus' surface and call $F(w)$ the complex flow potential associated with a set of $N$ vortices at positions $w_n$ with charges $q_n = \pm 1, \,n \in \{1,\dots,N\}$.\footnote{We neglect vortices with multiple elementary charge $\pm 2,\pm 3, \dots$ because they are expected to be energetically  unstable.} Close to each vortex core $w_n$ the flow field must behave just as in the flat plane, as described in Chapter IV of Ref.~\cite{Lamb1945book}:
\begin{align} 
	F(w) =  q_n \log (w-w_n) + \mO(w-w_n)\label{Condition_core}.
\end{align}
Additionally, the superfluid order parameter $\Psi$ must be periodic across the $\{u,v\}$ cell. This last condition means that the phase $\Phi=\Im \{F\}$ must wind in integer multiples $(k,m)$ of $2\pi$ upon traversing the cell in either direction: 
\begin{align}
\Im\{F(w+2\pi c)\} =& \Im\{F(w)\}+ 2\pi k,
\nonumber \\
\Im\{F(w+i\,2\pi b)\} =& \Im\{F(w)\}+ 2\pi m
. \label{Condition_boundary}
\end{align}

We can distinguish two classes of functions $F(w) $ fulfilling Eq.~\eqref{Condition_boundary}.
The first class contains simple linear functions of the form $ F_{k,m}(w)= w\left(i k/c + m/b \right)$ with integer coefficients $k,m$. These potentials represent uniform flows with $k$ elementary quanta of circulation around the toroidal direction, and $m$ quanta along the poloidal one. 
The second class consists of specific complex functions that are doubly periodic within the elementary cell, called \textit{elliptic functions} \cite{Whittaker1920Elliptic}. 
For our purposes, the following property is crucial: \textit{The sum of all residues of a doubly periodic function at its poles within any elementary cell is zero} (see Sec.~20.12 of Ref.~\cite{Whittaker1920Elliptic}). It follows from Eq.~\eqref{Condition_core} that the net vorticity in the system is zero: $\sum_n q_n = 0$.  In particular  the simplest possible configuration is a \textit{vortex dipole} with $\{q_1=1,q_2=-1\}$, which we consider in the next section.

\subsection{Vortex dipoles}

As seen in the bottom panel of Fig.~\ref{fig:ConformalMap}, the elementary  cell in the $\{u,v\}$ plane is a rectangle with periodic boundary conditions.   To satisfy these boundary conditions, imagine  tiling the whole plane with replicas of the original elementary cell.

For this purpose, it is convenient to consider a specific entire complex function 
\begin{align}
\vartheta_1(\zeta,p) =  2 p^{1/4} \sum_{n=0}^\infty (-1)^n p^{n(n+1)} \sin \left[\left(2 n+1\right)\zeta\right] \label{JacobiTheta} 
\end{align} 
known as the first Jacobi theta function~(see Chap.~21 of  \cite{Whittaker1920Elliptic}). It depends on the two dimensionless complex variables $\zeta$ and $p$ (with the restriction $|p|< 1$).  To ensure the latter constraint, we write $p = e^{i\pi \tau}$, where $\Im\{\tau\} > 0$. 
 In the complex $\zeta$ plane, the function $\vartheta_1(\zeta,p)$ has a periodic array of zeros at $\zeta_{km} = k\pi + m\pi \tau$, with $k,m \in \mathbb{Z}$.  
 Correspondingly $\log\vartheta_1(\zeta,p)$ has a periodic array of positive vortices at the same locations $\zeta_{km}$ [compare Eq.~(\ref{Condition_core})].

For the present rectangular cell, choose $\tau = i b/c$ to be pure imaginary, so that $p=\exp(-\pi b/c) < 1$.  The  function $\log \vartheta_1(\zeta - \zeta_1,p)$ then represents an infinite array of  positive vortices at $\zeta_1 +\zeta_{km}$.  In terms of the complex variable $w = u + iv$ with dimension of a length, consider a vortex dipole with positive vortex at $w_1$ and negative vortex at $w_2$ in the elementary cell.  It is now clear that what we call the ``classical'' complex potential for this vortex dipole 
\begin{align}
F_{\rm cl}(w)=\log\left[ \frac{\jt_1\left[(w-w_1)/2 c,p\right]}{\jt_1\left[(w-w_2)/2 c,p\right]}\right]\label{FlowPotential_Dipole_Classical}
\end{align}
indeed  represents the appropriate  periodic array of vortex dipoles.  

As shown below, however, $F_{\rm cl}(w)$   does not yield a periodic condensate wave function, similar to the case of a cylinder examined in  Ref.~\cite{Guenther2017}.
 This Jacobi theta function has the following  quasiperiodic properties $\jt_1(\zeta+\pi,p) = - \jt_1(\zeta,p) $ and $\jt_1(\zeta+\pi\tau,p) = - p^{-1} e^{-2 i \zeta} \jt_1(\zeta,p)$.  
 A detailed analysis shows that $\Im\{F_{\rm cl}\} \to \Im\{F_{\rm cl}\} + \Re\{w_{12}\}/c$ for $v \to v + 2\pi b$, namely one revolution along the $v$ axis, where   $w_{12}=w_1-w_2$.  This behavior violates Eq.~\eqref{Condition_boundary}.  An additional linear term ensures that the order parameter is indeed single valued:
\begin{align}
F(w) = \log\left[ \frac{\jt_1\left[(w-w_1)/2 c,p\right]}{\jt_1\left[(w-w_2)/2 c,p\right]}\right]-\frac{\Re\{w_{12}\}}{2 \pi b c} w.\label{FlowPotential_Dipole_Quantum}
\end{align}
This added linear term arises from the phase coherence of the superfluid across the entire torus' surface and is wholly quantum in nature.

The velocity field $\bw$ of the superfluid at a point $\bs(w)$ on the torus is given by Eqs.~\eqref{Phase_Flow} and \eqref{StreamFunction_Flow}, evaluated in the local tangent space of the surface at that point.
 Points at infinitesimally changed coordinates $\bs(w+dw)$ correspond to points at physical distances $|ds|=\lambda |dw|$ around it; specifically the basis vectors $1$ and $i$ in the complex plane $w= u+i v$ are equivalent to the set of local dimensionless tangent vectors $\bm u = \partial \bs/\partial u$ and $\bm v = \partial \bs/\partial v$, with norm $|\bm u|=|\bm v|=\lambda$. In this basis, the local gradient $\bm \nabla$ is $\bm \nabla = \tilde{\bm \nabla}/\lambda$, where 
\begin{align}\label{Gradient_NaturalBasis}
\tilde{\bm \nabla} = \bm u \ \partial_u +  \bm v \ \partial_v.   \tilde{\bm \nabla},
\end{align}

Equations \eqref{Phase_Flow} and \eqref{StreamFunction_Flow} then become 
\begin{align}
{\bf w} =& \frac{\hbar}{M \lambda} \tilde{\bm \nabla} \Phi(\bs) \\
{\bf w} =& \frac{\hbar}{M \lambda} \hat{\bn}\times \tilde{\bm \nabla } \chi(\bs),
\end{align}
which can  be combined in the complex form
\begin{align}
\Omega = \mathbf{w}_v + i \mathbf{w}_u = \frac{\hbar}{M} \frac{F'(w)}{\lambda}, \label{PotentialFlow_Curved}
\end{align}
analogous to Eq.~\eqref{PotentialFlow}. Here $\Omega$ denotes the complex representation of the velocity field $\mathbf{w}$. Figure~\ref{fig:FlowPotential_Dipole} gives an example of a vortex-dipole configuration and the corresponding phase pattern, streamlines and flow field,
both in the $\{u,v\}$ coordinate set (top panels) and mapped to the surface of a torus (bottom panels).

By construction, the imaginary part of Eq.~(\ref{FlowPotential_Dipole_Quantum}) is doubly periodic, as is evident from the colors in the upper left panel of Fig.~\ref{fig:FlowPotential_Dipole}.  In contrast, the real part is only quasiperiodic: 
\begin{equation}
F(w+2\pi c k +i \, 2\pi  b m)
=F(w)- \ k \frac{\Re\{w_{12}\}}{b} - \ m \frac{\Im\{w_{12}\}}{c},
\end{equation}
with integers $k$ and $m$.
Nevertheless, the physical velocity field in Eq.~(\ref{PotentialFlow_Curved}) is doubly periodic, as is clear from its representation in terms of the gradient of the  doubly periodic phase function $\Phi = \Im\{F\}$.
Since $\chi=\Re \{F \}$, streamlines $\chi=c_1$ that  extend to the boundary of the $\{u,v\}$ cell must be identified with other streamlines $\chi=c_1+k \Re \{w_{12}\}/b + m \Im\{w_{12}\}/c$, with $k,m \in \mathbb{Z}$. 
Hence, these curves might wind several times around the torus before closing on themselves, as can be seen in the lower left panel of Fig.~\ref{fig:FlowPotential_Dipole}.

In the limit of large major radius,  $a/b \gg 1$, we expect the torus to approximate the simpler geometry of an infinite cylinder with radius $b$. The theta function Eq.~\eqref{JacobiTheta} has the remarkable imaginary transformation  
$(-i\tau)^{1/2}\jt_1(\zeta,p) = -i  \exp(i \tau' \zeta^2/\pi) \jt_1(\zeta \tau',p')$, with $\tau'= -1/\tau$ and $p'= \exp(i\pi \tau') = e^{-\pi c/b}$. 
In the infinite-cylinder limit, where $c/b \rightarrow \infty$, one finds $p \rightarrow 1$ and correspondingly $p'\rightarrow 0$.
The series Eq.~\eqref{JacobiTheta} for the transformed theta function with parameter  $p'$ now terminates after the first term and gives
\begin{align}
F_{\rm cyl}(w) = \log\left[\frac{\sin[i(w-w_1)/2 b]}{\sin[i(w-w_2)/2b]}\right]. \label{LimitCylinder}
\end{align}
This expression agrees with the flow potential of a vortex dipole on an infinite cylinder of radius $b$ that  we derived in Ref.~\cite{Guenther2017}, provided one identifies $u$ with the coordinate along the axis of the cylinder, and $v$ with the azimuthal coordinate around its radius. 
Note that the linear term in Eq.~(\ref{FlowPotential_Dipole_Quantum}) disappears in this limit.

\subsection{Flux quanta on the torus and ``winding'' of single vortices}

The function $F(w)$ also has quasiperiodic properties in terms of the  positions $w_{1},w_2$ of the two vortices. For example, consider $w_1\rightarrow w_1 + 2\pi c \, k + i \,2 \pi  b \, m$ with $k,m$ integer. 
Using the properties of the Jacobi theta function we find
\begin{align}
F(w) \rightarrow F(w) + q_1\left(\frac{m}{c}+ i \frac{n}{b}\right)w. \label{FluxQuanta}
\end{align}
Winding a single vortex
once in the toroidal direction along $u$ with $n=1$ and $m=0$   introduces an elementary quantum of flux around the poloidal direction along $v$, and vice versa.  When $w_1=w_2$, the complex potential defined in Eq.~\eqref{FlowPotential_Dipole_Quantum} reduces 
 to $F(w)=0$.  According to Eq.~\eqref{FluxQuanta},  if one creates an infinitesimal vortex-antivortex pair and then annihilates it after ``winding'' one of its members  around the torus, the superfluid initially at rest acquires an  elementary flux quantum for flow in the orthogonal direction around the torus. When a poloidal winding generates a flux in the toroidal direction, the two-dimensional vortex dipole can be considered the entry and exit point of a single flux line crossing into the inside of the torus. This model realizes the 
classic picture of phase slippage in ring-shaped superfluids, as  discussed by Anderson in Ref.~\cite{Anderson1966}.

\subsection{Multipole configurations}

The extension to cases with more than two vortices is straightforward. Defining 
\begin{align}
F(w,w_n) = \log\left[\jt_1\left(\frac{w-w_n}{2c},p\right)\right]-\frac{\Re\{w_n\}}{2 \pi b c} w, \label{Potential_Monopole}
\end{align}
we can write Eq.~\eqref{FlowPotential_Dipole_Quantum} as
\begin{align}
F(w) = \sum_n q_n F(w,w_n). \label{FlowPotential_Multipole}
\end{align}
Every neutral set of $2N$  vortices with elementary charges $q_n=\pm 1$ can be organized into a set of $N$ dipoles.  In this way, Eq.~\eqref{FlowPotential_Multipole} describes every possible neutral set of vortices on a torus. Correspondingly, we  define $f(w,w_n)=\partial_w F(w,w_n)$, which explicitly gives 
\begin{align}
f(w,w_n) = \frac{1}{2 c} \frac{\jt_1'\left[(w-w_n)/2c,p\right]}{\jt_1\left[(w-w_n)/2c,p\right]}- \frac{\Re\{w_n\}}{2\pi b c}.\label{Velocity_Monopole}
\end{align}
The complex velocity field $\Omega(w)$ then becomes 
\begin{align}
\Omega(w) = \frac{\hbar}{M}\frac{1}{\lambda} \sum_n q_n f(w,w_n). \label{FlowField_Multipole}
\end{align}
An individual  term $q_n F(w,w_n)$ in Eq.~(\ref{FlowPotential_Multipole}) can be interpreted as the complex potential of a single vortex on the torus, but this picture applies only for an ensemble of vortices with zero net charge.


\section{Energy of vortex configurations}\label{sec:Energy}

\begin{figure}[t]
	\begin{centering}
		\includegraphics[width=0.3\textwidth]{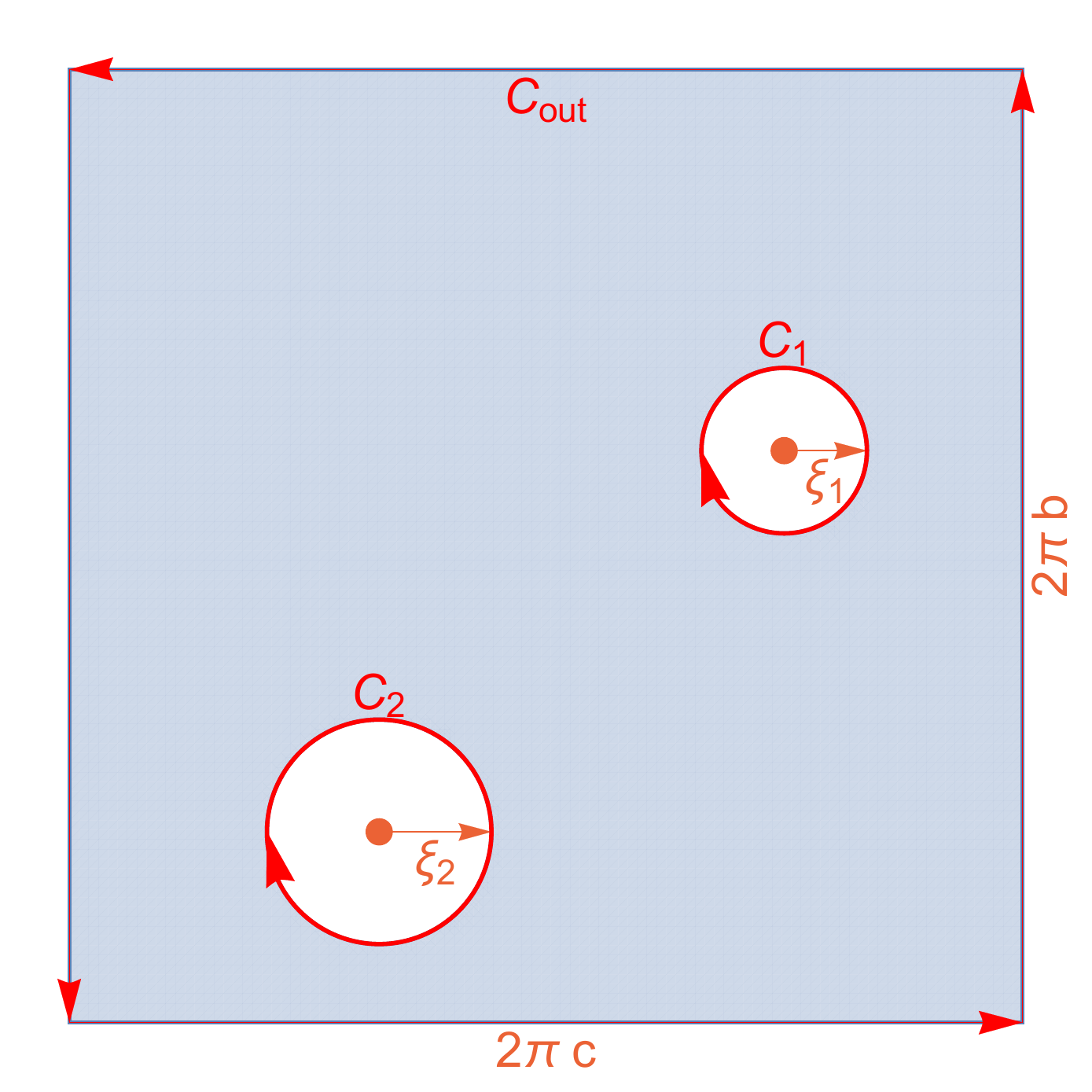}
	\end{centering}
	\caption{Region of integration $A$ in the $\{u,v\}$ plane for the surface integral in Eq.~\eqref{Energy_Kinetic} determining the kinetic energy for a vortex dipole on a torus. 
		On the torus' surface, a small disk of physical size $\xi$ around each vortex center must be excluded from the area of integration.  In the curvilinear $\{u,v\}$ plane, this physical  size
		is locally rescaled with
		the  factor $\lambda(v)= c/[a-b\cos(v/b)]$.  As a result, 
		the excluded disk now has 
		a  radius $\xi_n = \xi/\lambda(v_n)$, where $v_n$ is the $v$-coordinate of the $n$th vortex core.  Green's theorem in two dimensions leads to a corresponding boundary integral (red arrows)  
		that   can be 
		separated into two small circles $C_{1,2}$ and one outer boundary $C_{\text{out}}$.
	}
	\label{fig:Energy_Area_Contour}
\end{figure}

The energy of a vortex configuration is given by the kinetic energy of the flow field
\begin{align}
E = \frac{1}{2} M  \rho \int_A du \,dv  \, \lambda^2 |\mathbf{w}|^2, \label{Energy_Kinetic}
\end{align}
with $A$ the appropriate integration region. Before proceeding, 
we need to define carefully   the region $A$.

On first impression, $A$ denotes the $\{u,v\}$ elementary cell $\{ -\pi c, \pi c \} \times \{-\pi b, \pi b\}$, but 
close to each vortex core $w_n$, the squared velocity $|\mathbf{ w}|^2\sim 1/|w-w_n|^2$ diverges.  As a result, the integral  must exclude  a small disk 
around each core center  to 
reflect the finite \textit{physical} core size $\xi$ on the torus.  As long as $\xi$ is much smaller than the smallest geometrical scale of the system, the vortex is still
effectively point-like,
and the torus' surface can be considered locally flat across the 
vortex core. In the $\{u,v\}$ plane, however,  the physical length scale $\xi$ appears  locally rescaled. 
Specifically, the original disk of physical radius $\xi$ on the torus becomes  a disk with radius $\xi_n = \xi/\lambda_n$ in $\{u,v\}$ coordinates, where $\lambda_n=\lambda(u_n,v_n)$ is the scale factor evaluated at the vortex core. 
Figure~\ref{fig:Energy_Area_Contour} illustrates this transformation  for a vortex dipole.

In the curvilinear $\{u,v\}$ coordinates, the earlier Eq.~(\ref{StreamFunction_Flow}) acquires a scaling factor: 
$\mathbf{w}= \hbar/ (M \lambda) \, \hat{\bm n} \times\bm \tilde{\bm \nabla} \chi$, where $\tilde{ \bm \nabla}$ is defined in Eq.~\eqref{Gradient_NaturalBasis}.  Consequently, the scale factors in Eq.~(\ref{Energy_Kinetic}) cancel and we find
\begin{align}
E =& \frac{\rho \hbar^2}{2 M} \int_A du \ dv \ (\tilde{\bm \nabla} \chi)\cdot(\tilde{\bm \nabla} \chi) \nonumber \\
=& \frac{\rho \hbar^2}{2 M} \int_A du \ dv \ \left[\tilde{\bm  \nabla} \cdot \left(\chi \tilde{\bm \nabla} \chi \right)- \chi \tilde{\nabla}^2 \chi \right] \nonumber \\
=& \frac{\rho \hbar^2 }{2 M} \left(\int_{\partial A} d {\bm s}\times\hat{\bm n} \cdot \chi \tilde{\bm \nabla} \chi-  \int_A du \ dv \ \chi \tilde{\nabla}^2 \chi\right), \label{Energy_Stream_terms}
\end{align}
where we used integration by parts in the first step and Green's theorem    (essentially the divergence theorem) in two dimensions  in the second.  We can write $\chi = \sum_n q_n \chi_n$ with $\chi_n= \Re \{ F(w,w_n) \}$. It is straightforward to confirm that $\tilde{\nabla}^2 \chi = 2\pi \sum_n q_n \delta(u-u_n)\delta(v-v_n)$. Dirac delta function. 
Hence the second term in Eq.~(\ref{Energy_Stream_terms}) vanishes because the region $A$ excludes  all vortex centers.

The  first term  in Eq.~(\ref{Energy_Stream_terms}) is a boundary 
integral that separates into a counter-clockwise (positive) loop around the outer boundary of the elementary cell and clockwise (negative) loops of radius $\xi_n$ around each vortex center $w_n$.  
The integral around the outer boundary of 
the elementary cell vanishes because of  the quasiperiodic properties of the complex potential in Eq.~\eqref{FlowPotential_Multipole}.
The remaining parts are the small circles
$C_n$ around each vortex core:
\begin{align}
E =& \frac{\hbar^2 \rho}{2 M} \sum_{n} \int_{C_n} d {\bm s}\times \hat{\bm n} \cdot \chi\tilde{ \bm \nabla}\chi\nonumber \\
=& \frac{\hbar^2 \rho}{2 M} \sum_{n,m,k} q_m q_k \int_{C_n} d {\bm s}\times \hat{\bm n}  \cdot \chi_m \tilde{\bm \nabla} \chi_k, \label{Energy_Terms_Circles}
\end{align}
where $\tilde{\bm \nabla}$ acts on the first variable of $\chi_k = \Re\{F(w,w_k)\}$

Near the $n$th vortex, it is convenient to introduce the  local vector $\bm\delta = (u-u_n)\,\bm u + (v-v_n)\,\bm v$, of length $|\bm \delta|= \xi$, for points $w$ on $C_n$.  Correspondingly, the local stream function behaves like $\chi_n \approx \log \delta $ apart from an additional constant. 
Most terms appearing in Eq.~\eqref{Energy_Terms_Circles} will be negligible 
because of the small length $2\pi \xi$ of each curve $C_n$. The dominant contributions come from terms with $k = n$, since the singular behavior of
$\tilde{\bm\nabla }\chi_n \approx \hat{\bm \delta}/\delta$ cancels the small circumference on $C_n$.

For $m \neq n$,  the function $\chi_m$ reduces to the constant  
\begin{equation}
\chi_{nm} = \Re\{F(w_n,w_m)\}, 
\end{equation}
apart from small corrections of order $\xi/b$.
Terms with $m = n$ require special care, however, because of the logarithmic behavior of $\chi_n\approx \log\delta$. 
On the curve $C_n$, we can set $\bm \delta = \xi_n \cos \theta \ \bm  u + \xi_n \sin \theta \ \bm v $ and define
\begin{align}
\chi_{nn}= \log\left(\frac{\jt_1' \xi_n}{2c}\right)- \frac{\Re\{w_n\}^2}{2 \pi b c}, \label{Self_Energy}
\end{align} 
where $\jt_1' = \partial\jt_1(\zeta,p)/\partial \zeta\big|_{\zeta=0}$ 
 is a constant. We can then write the dominant contributions in Eq.~\eqref{Energy_Terms_Circles} in the compact form 
\begin{align}
\frac{2 M E}{\hbar^2 \rho}= - 2\pi \sum_{n,m} q_n q_m \chi_{nm} + \mO(\xi/b) \label{Energy_Final}.
\end{align}

Note that this expression   involving a double sum over  $\chi_{nm}$ takes the same form as that for a set of vortices in the plane. Omitting the constant core-energy term in Eq.~\eqref{Self_Energy}, we can group the terms in Eq.~\eqref{Energy_Final} into three categories according to their physical origin:
\begin{align}\label{Energy_Terms}
\frac{2 M E}{\rho \hbar^2} =&E_{\rm class} + E_{\rm curv}+E_{\rm quant}\\
=&-2\pi {\sum_{n,m}}^{'} q_n q_m \log\left| \jt_1\left(\frac{w_n-w_m}{2 c},p \right)\right|\nonumber\\
&+ 2\pi \sum_n \log\lambda_n\nonumber\\
&+ \sum_{n,m} q_n q_m \frac{u_n u_m}{b c},\nonumber
\end{align}
where the primed double sum omits terms with $n = m$.
The {\it classical} term $E_{\rm class}$ gives the energy of point vortices in a classical inviscid, incompressible fluid contained in the elementary cell with a constant metric, which we may call a ``flat'' torus.
The {\it curvature} term $E_{\rm curv}$ represents changes in the kinetic energy due to the curvature of the torus induced by the locally distorted metric Eq.~\eqref{LineElement_uv} and would not be present for a flat metric. In the context of simply connected two-dimensional superfluids, these terms have been discussed extensively in \cite{Turner2010}.
Finally, as discussed in the previous section, the {\it quantum} term $E_{\rm quant}$ arises from 
the need to ensure that  the superfluid order parameter is single valued and hence represents a purely quantum contribution.  The next section shows how
these terms generate distinct contributions to the vortex-core dynamics. 

\begin{figure*}[t!]
	\begin{centering}
		\subfloat[][	\label{fig:Dipole_Trajectories:11}]{	\includegraphics[width=0.245\textwidth]{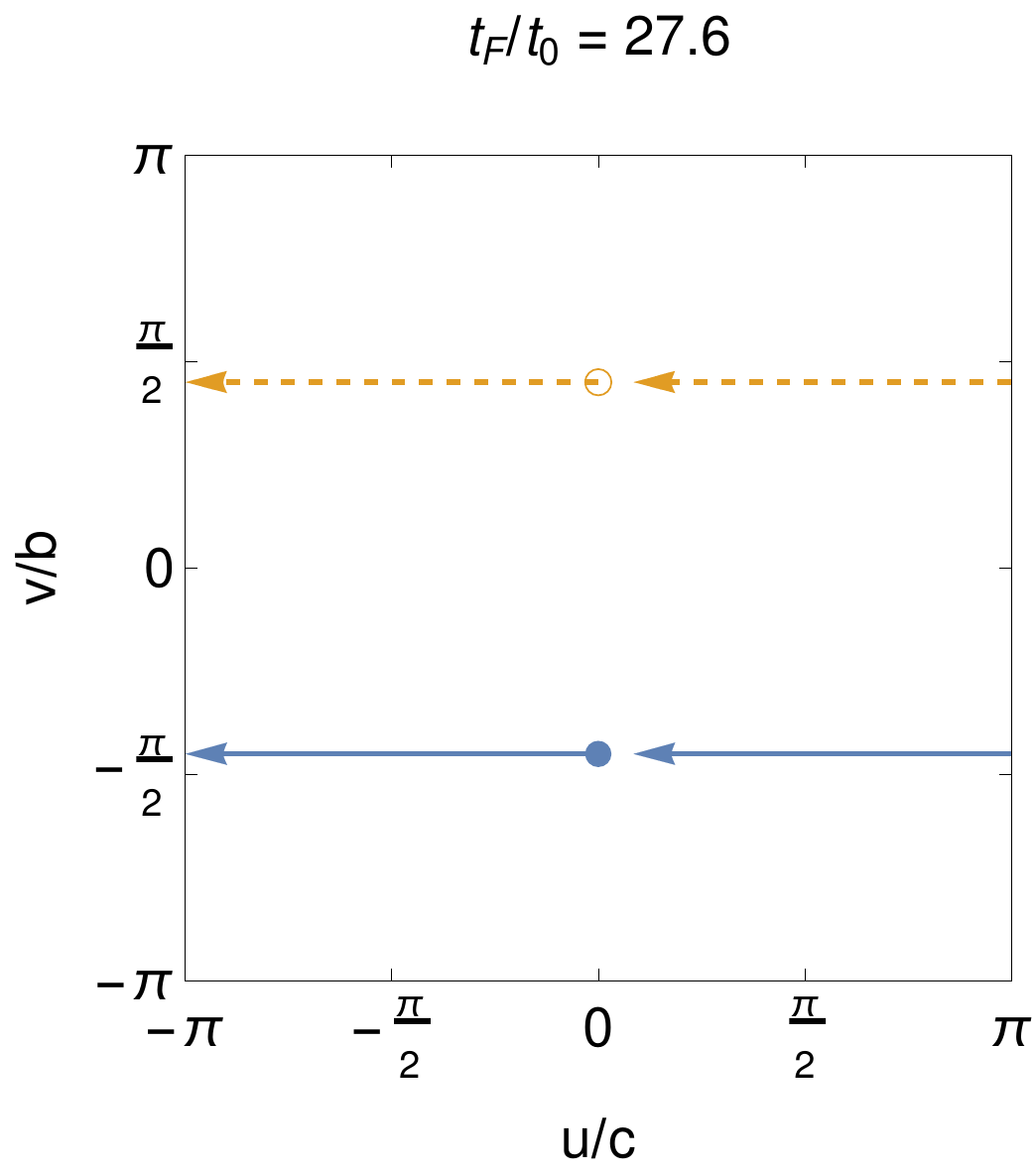}}
		\subfloat[][	\label{fig:Dipole_Trajectories:12}]{	\includegraphics[width=0.245\textwidth]{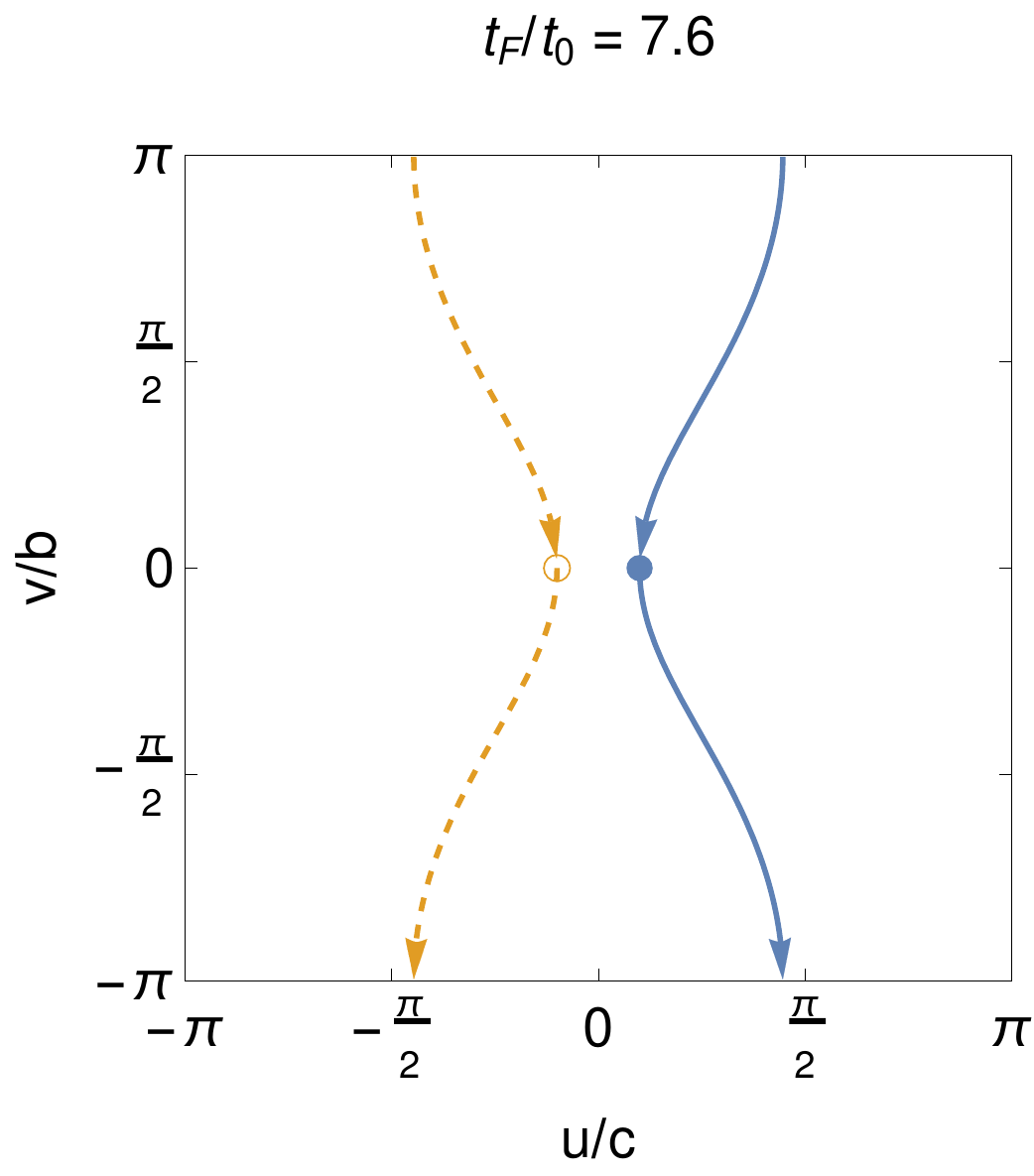}} 
		\subfloat[][	\label{fig:Dipole_Trajectories:13}]{	\includegraphics[width=0.245\textwidth]{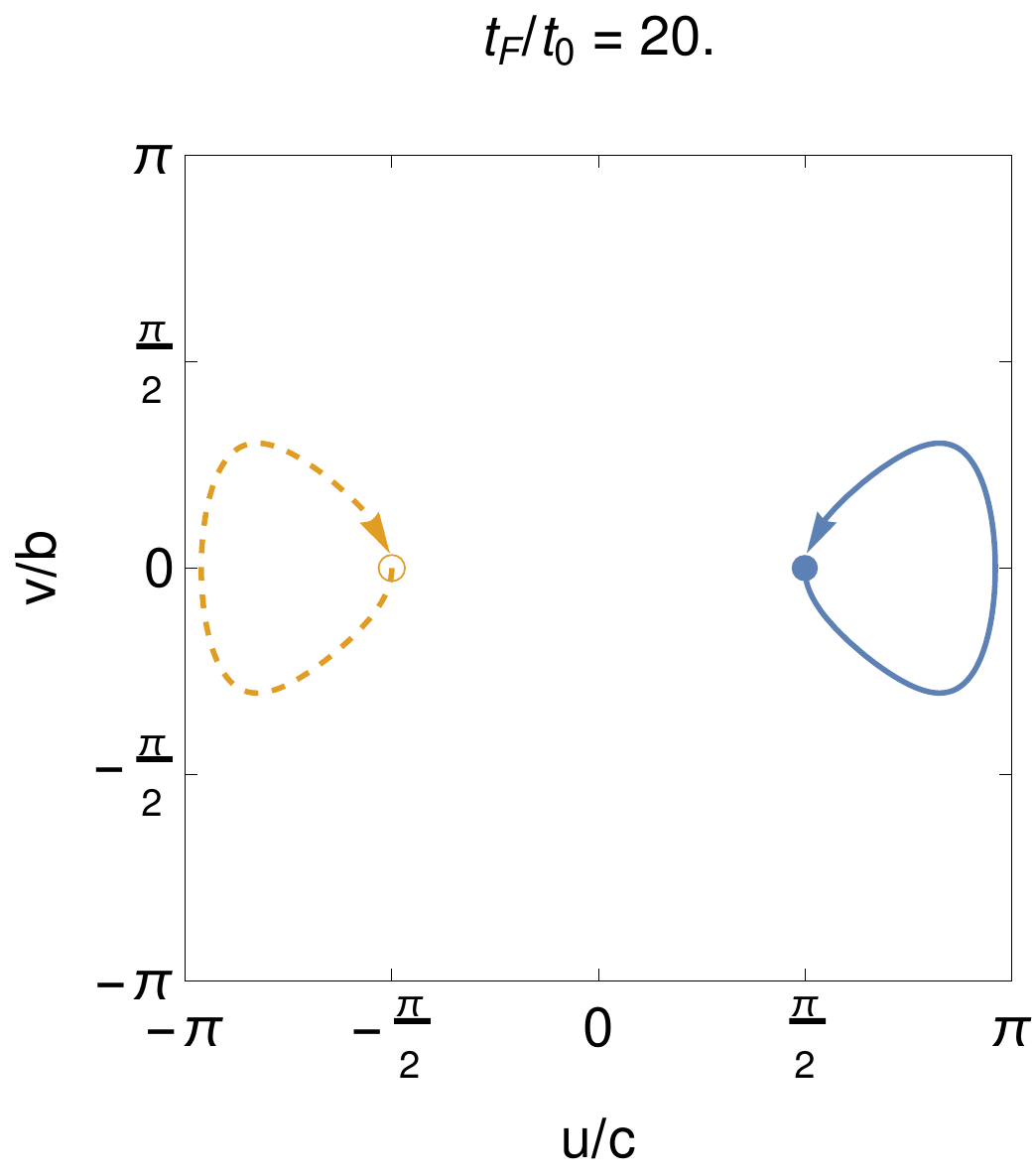}}
		\\
		\subfloat[][	\label{fig:Dipole_Trajectories:21}]{	\includegraphics[width=0.245\textwidth]{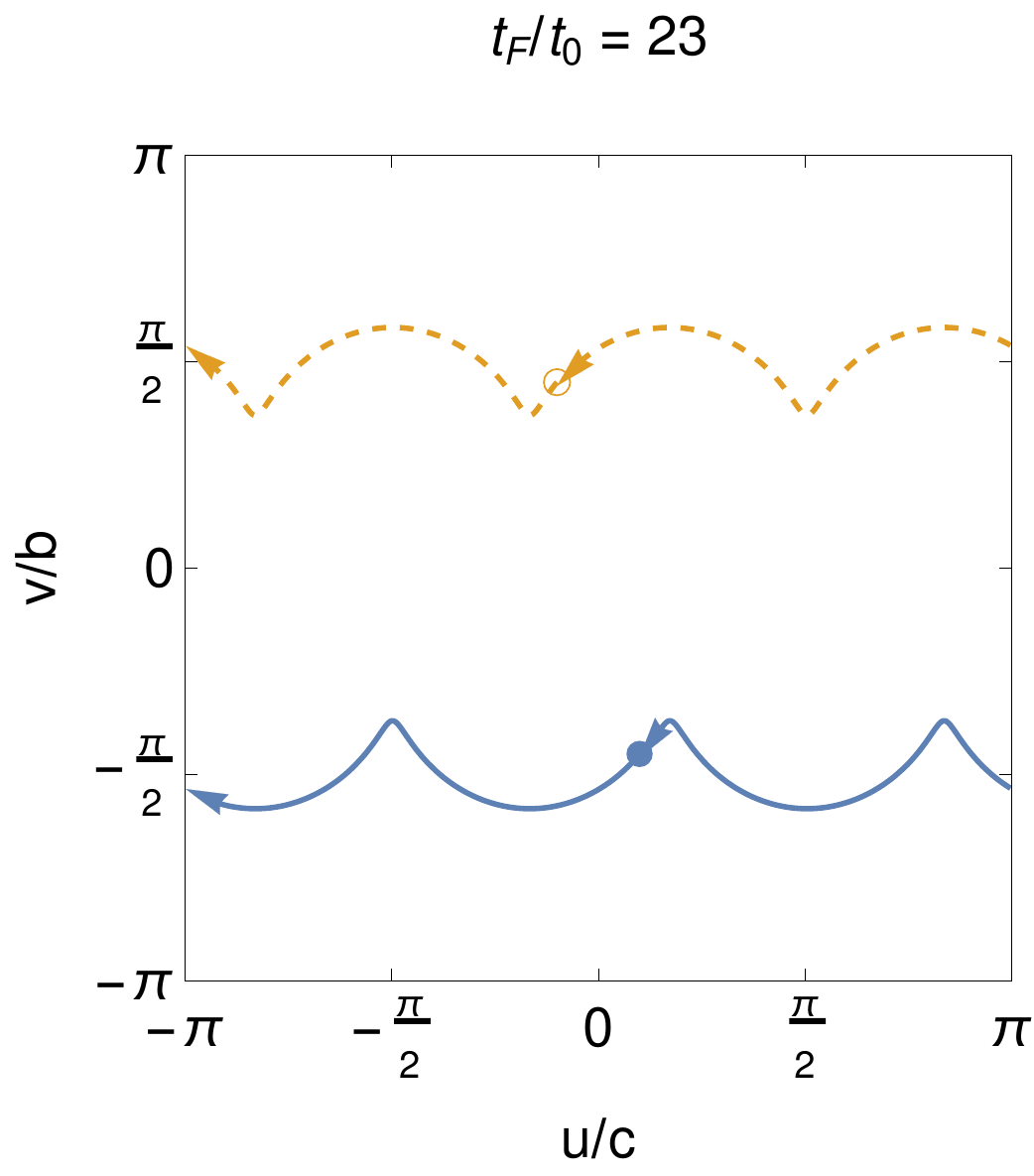}}
		\subfloat[][	\label{fig:Dipole_Trajectories:22}]{	\includegraphics[width=0.245\textwidth]{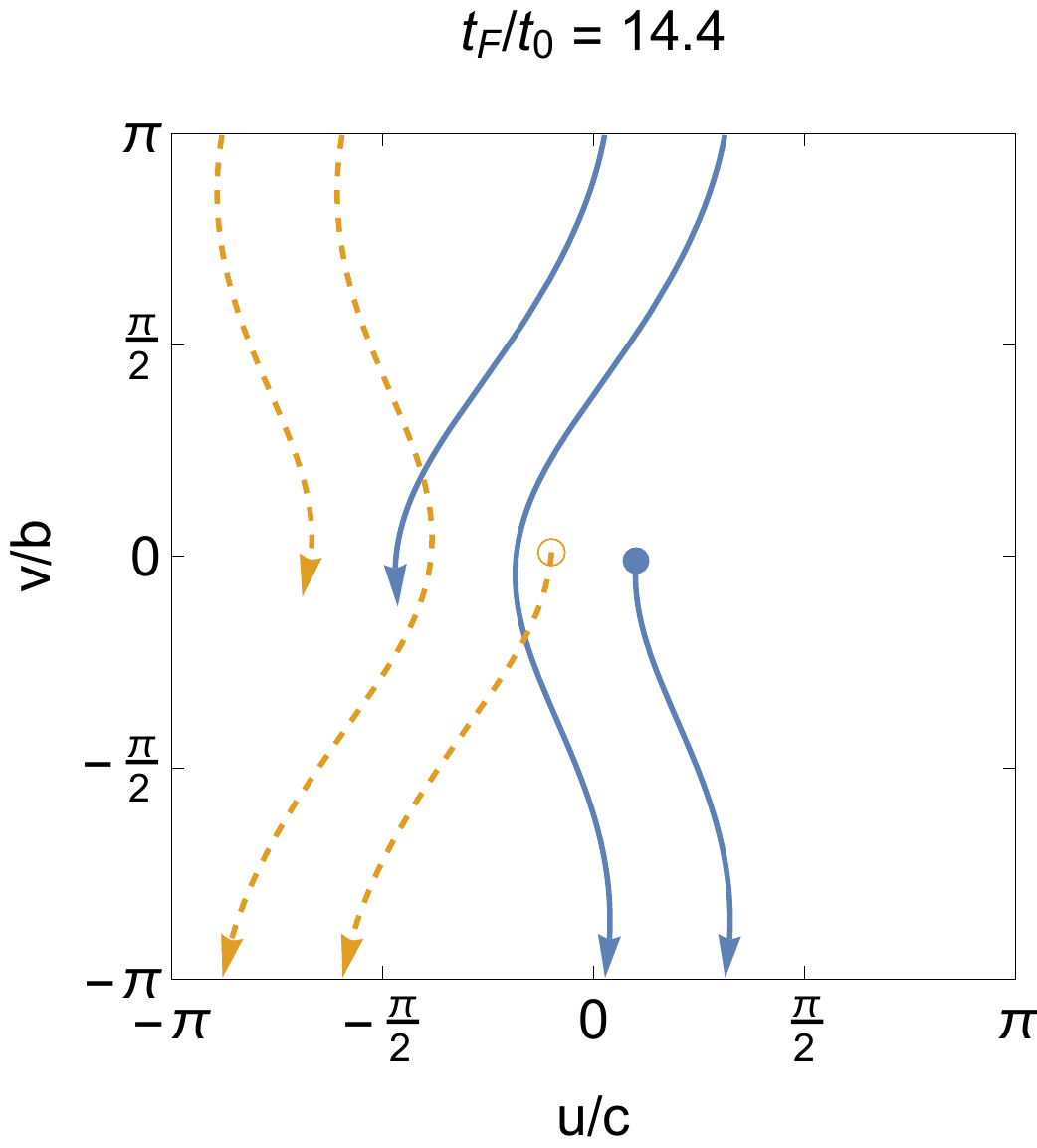}}
		\subfloat[][	\label{fig:Dipole_Trajectories:23}]{	\includegraphics[width=0.245\textwidth]{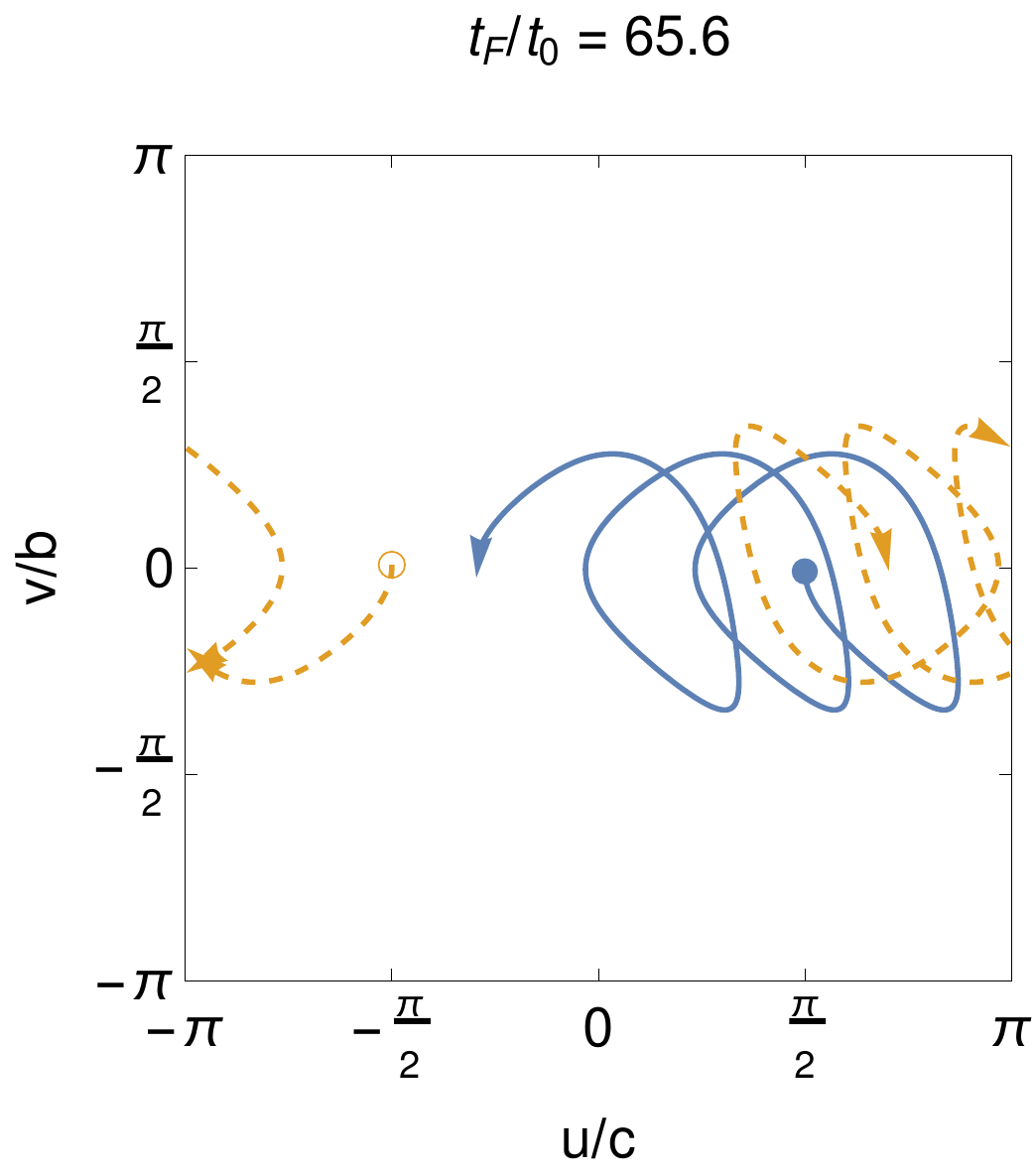}}
		\caption{
			Trajectories of a vortex dipole on the surface of 3D torus shown in the $u,v$ plane. Initially a vortex is set at position $z_{1,0}$(blue dot) and an anti-vortex at position $z_{2,0}$(orange dot), their trajectories (blue and orange arrows respectively) are then integrated according to Eq.~\eqref{Hamiltonian_Vortices}, up to a final time $t_F$, measured in units $t_0 = M b^2/\hbar$. \textit{Symmetric toroidal configuration:} The vortex dipole moves rigidly in a straight line perpendicular to the dipole axis (\ref{fig:Dipole_Trajectories:11}). \textit{Symmetric poloidal configuration:} The curvature and quantum terms in Eq.~\eqref{Energy_Terms} initially drive the dipole apart  in circling the torus in the poloidal direction (\ref{fig:Dipole_Trajectories:12}).  For larger separations, they alter the motion   into  closed loops on the outer side of the torus (\ref{fig:Dipole_Trajectories:13}). \textit{Asymmetric configurations:} dipoles not aligned symmetrically around either axis experience a hybrid trajectory. 
			Configurations close to the symmetric cases lead to a combination of both types of trajectories on separated time scales, as shown in (\ref{fig:Dipole_Trajectories:21}-\ref{fig:Dipole_Trajectories:23}).
		}
		\label{fig:Dipole_Trajectories}
	\end{centering}	
\end{figure*}


\section{Vortex dynamics}\label{sec:Dynamics}

The time-dependent evolution of the flow field of a two-dimensional inviscid, incompressible fluid containing point vortices can be reduced entirely to the motion of the vortex cores, as first established through Helmholtz's famous vorticity theorems \cite{Helmholtz1858}. Each vortex moves 
with the fluid at the ``local'' velocity $\bw_n$  at its core position, regularized to 
omit the divergent contribution coming from the vortex itself.  A detailed calculation in the Appendix gives [see Eq.~(\ref{Velocity_Curved_Final})]
\begin{align}
\mathbf{w}_n
=& \frac{\hbar}{M} \frac{1}{\lambda(\bs_n)} \hat{\bn} \times \tilde{\bm \nabla} \left(\chi_n^{\rm reg}(\bs)-\frac{q_n}{2}\log\lambda(\bs)\right)_{\bs \rightarrow \bs_n} \label{Velocity_Curved}
\end{align} 
for any isothermal coordinate set $\bm s = \{u,v\}$ with a general scaling factor $\lambda(\bm s)$.  Here 
\begin{equation}
\chi_n^{\rm reg}(\bm s) = \Re\{F(w) - q_n\log(w-w_n)\}
\end{equation}
is the regular part of the stream function near $w_n$.

  Let  $\Omega_n = \bw_{n,v}+i \bw_{n,u}$ be the complex velocity of the $n$th vortex core. 
For a set of $N$ vortices on a torus,
the Appendix shows that  Eq.~(\ref{Velocity_Curved})  has the equivalent complex  form  
\begin{align}
\Omega_n = \frac{\hbar}{M}\frac{1}{ \lambda(v_n)}\bigg(& \sum_{m\neq n}q_m f(w_n,w_m)   \nonumber \\
 &  + i \frac{q_n}{2} \frac{\lambda'(v_n)}{\lambda(v_n)} - \ q_n \frac{u_n}{2 \pi b c} \bigg)
 \label{Velocity_Vortex_Torus},
\end{align} 
with $f(w_n,w_m)$ defined by Eq.~\eqref{Velocity_Monopole}. 

Note that Eq.~\eqref{Velocity_Vortex_Torus} is  the physical velocity of the vortex.  In the curvilinear coordinate set $\bm s = \{u,v\}$, however,  the equations of motion for each vortex take the form $\dot{u}_n = \Im\{\Omega_n\}/\lambda_n$ and $\dot{v}_n = \Re\{\Omega_n\}/\lambda_n$, where $\lambda_n = \lambda(\bm s_n)$.

 Kirchhoff~\cite{Kirchhoff1876} introduced the Hamiltonian formulation for classical point vortices on a plane
 (see also Sec.~157 of \cite{Lamb1945book}) and Refs.~\cite{Lin1941-1,Lin1941-2} extended the description to include more general planar geometries.  Here, a detailed analysis shows 
 that  vortex dynamics on a torus can  be recast in a 
 Hamiltonian form, 
where the coordinates $u_n$ and $v_n$ serve as  
conjugate variables and Eq.~\eqref{Energy_Final} gives the energy function: 
\begin{align}
\rho \hbar \, 2\pi q_n \, \dot{u}_n =& \frac{1}{\lambda_n^2} \frac{\partial}{\partial v_n} E(\bm s_1,\dots,\bm s_N) \nonumber \\
\rho \hbar \, 2\pi q_n \, \dot{v}_n =& -\frac{1}{\lambda_n^2} \frac{\partial}{\partial u_n} E(\bm s_1,\dots,\bm s_N) \label{Hamiltonian_Vortices},
\end{align}  
This formulation ensures 
 that the time evolution conserves the energy in  Eq.~\eqref{Energy_Final}, as it must because the system has no dissipation.

The first set of terms in Eq.~\eqref{Velocity_Vortex_Torus} describes the velocity induced by all other vortices at the position of the $n$th vortex core. 
The second term corresponds to the curvature term $E_{\rm curv}$ in  Eq.~\eqref{Energy_Terms};  it induces a purely toroidal motion.
The last term is the effect of the quantum term $E_{\rm quant}$ in  Eq.~\eqref{Energy_Terms} of the $n$th vortex on itself, inducing a purely poloidal motion. 

In a flat two-dimensional plane, vortex dipoles always move rigidly in the direction perpendicular to the dipole axis given by the right-hand rule. On the torus, however, the possible trajectories are much more complicated, as depicted in Fig.~\ref{fig:Dipole_Trajectories}, where Eq.~\eqref{Hamiltonian_Vortices}  was integrated numerically for a representative sample of initial vortex positions. The ``classical'' rigid dipole motion occurs only  for a ``poloidal'' initial vortex dipole aligned along the poloidal axis  and symmetric with respect to the $v=0$ line in the $\{u,v\}$ plane as shown in Fig.~\ref{fig:Dipole_Trajectories:11}. In this case, the motion is purely toroidal, analogous to that on an infinite cylinder. In contrast, the poloidal movement of a toroidal dipole induces stretching of the toroidal distance between vortices  (Fig.~\ref{fig:Dipole_Trajectories:12}) due to the $v$ dependence of the curvature terms Eq.~\eqref{Energy_Terms}.  Indeed, the vortices can even execute counter-rotating loops (Fig.~\ref{fig:Dipole_Trajectories:13}) for sufficiently large initial separation.  The remainder of Fig.~\ref{fig:Dipole_Trajectories} illustrates various other initial configurations.

Of interest are also static dipole configurations, where $\Omega_{1,2} = 0$. Symmetric configurations with either a purely toroidal or poloidal dipole axis allow for such solutions, depending on the geometric scales $a$ and $b$.  Interestingly, for a ``classical'' fluid without the $E_{\rm quant}$ term, a toroidal dipole would be static in the case of 
 $w_{1} =w_2 = \pm \pi c/2$, when the vortex cores are separated by the maximal distance possible on the torus. The quantum term $E_{\rm quant}$  however forces this configuration to move (see Fig.~\ref{fig:Dipole_Trajectories:13}) and the equilibrium distance varies. 

\section{Generalized toroidal surfaces of revolution}\label{sec:GenralizedRS}

The discussion above can be extended to surfaces similar to the torus by finding an appropriate isothermal coordinate set where the complex potential formalism applies.
Consider a two-dimensional curve, described parametrically as ${\bm {f}}(\eta)=\{f_1(\eta),f_2(\eta)\}, \  \eta \in \{-\pi,\pi\}$, such that $f$ is closed and never crosses itself. A corresponding surface of revolution has the parametrization  [compare Eq.~(\ref{TorusParametrization})]
\begin{align}
\bs(\phi,\eta) = \begin{cases}
x = [a+f_1(\eta)] \cos \phi \\
y =  [a+f_1(\eta)] \sin \phi \\
z = f_2(\eta)
\end{cases} \qquad \phi,\eta \in \{-\pi,\pi \}, \label{Revolution_Parametrization}
\end{align}
where we restrict to cases $a+\min_{\eta}f_1(\eta) > 0$, such that the resulting surface has a central hole and hence is multiply connected. Such surfaces are known as toroids. The line element corresponding to Eq.~\eqref{Revolution_Parametrization} is
\begin{align}
ds^2 = \underbrace{\left[a+ f_1(\eta)\right]^2}_{\lambda_\phi(\eta)^2} d\phi^2 + \underbrace{\left[f_1'(\eta)^2+f_2'(\eta)^2\right]}_{\lambda_\eta(\eta)^2} d\eta^2, \label{LineElement_RS}
\end{align}
defining $\lambda_\phi$ and $\lambda_\eta$. 
When ${\bm {f}}$ 
describes a circle with radius $b$, these expressions yield 
the parametrization  of a torus in Eq.~\eqref{TorusParametrization}.

\subsection{Isothermal coordinate set}

\begin{figure*}[t!]
	\begin{centering}
		\includegraphics[width=0.15\textwidth]{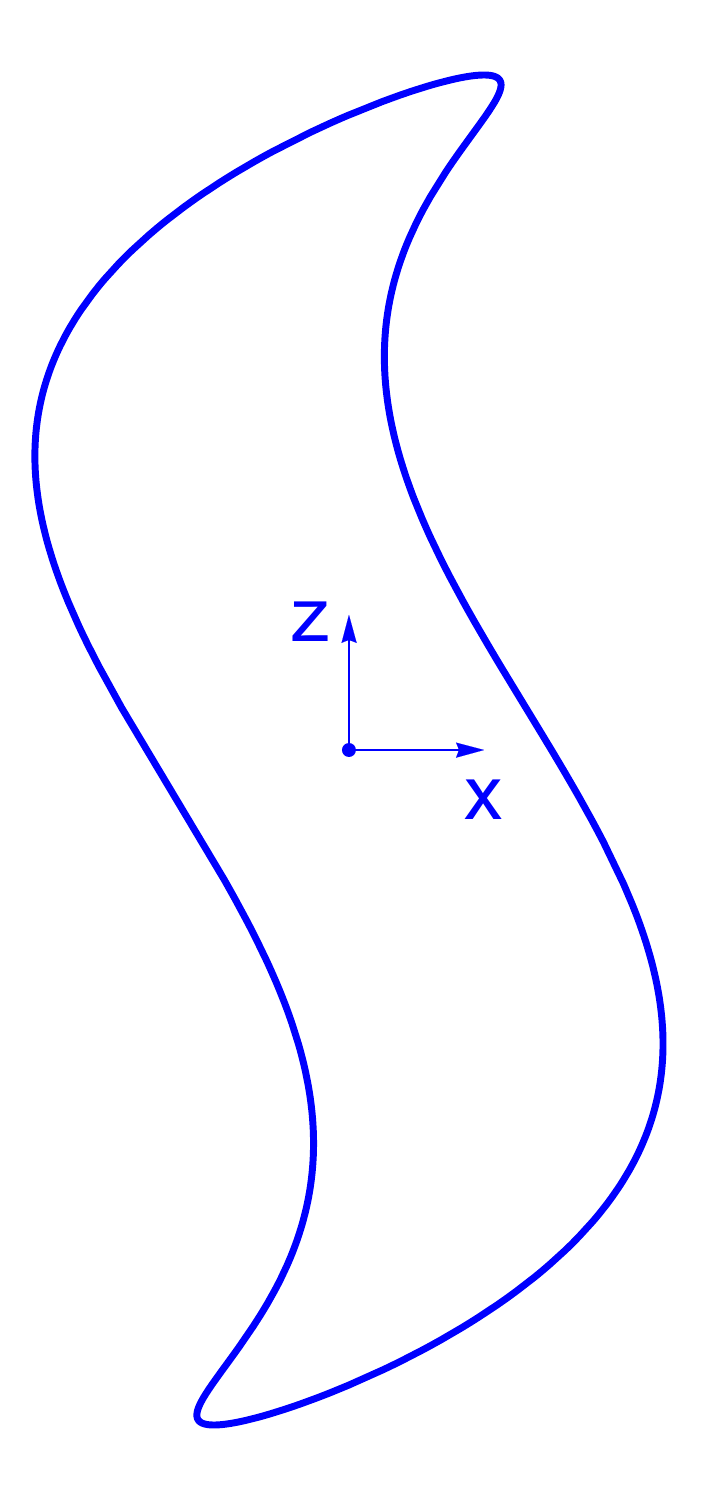}
		\hspace{20pt}
		\includegraphics[width=0.28\textwidth]{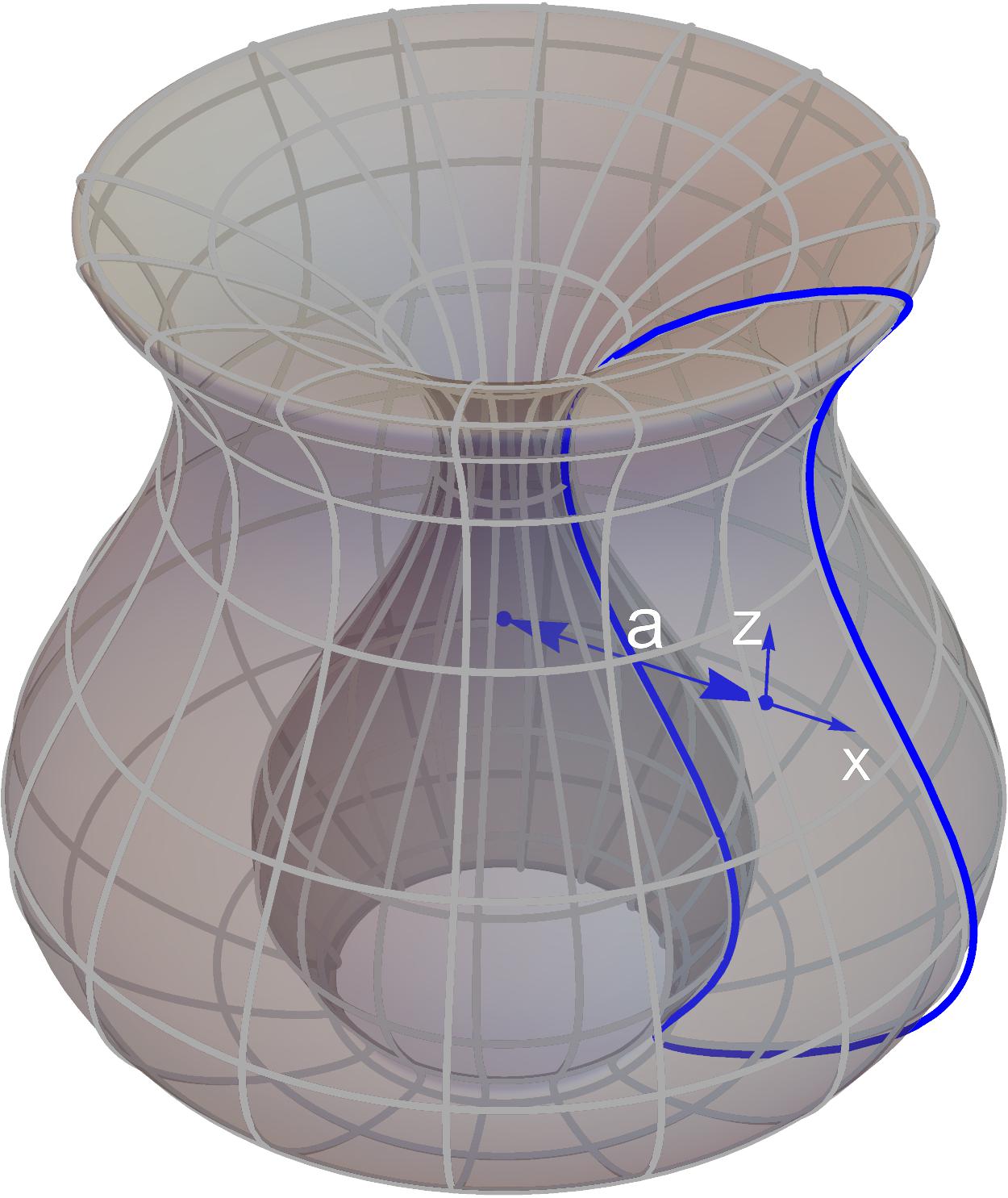}
		\hspace{10pt}
		\includegraphics[width=0.28\textwidth]{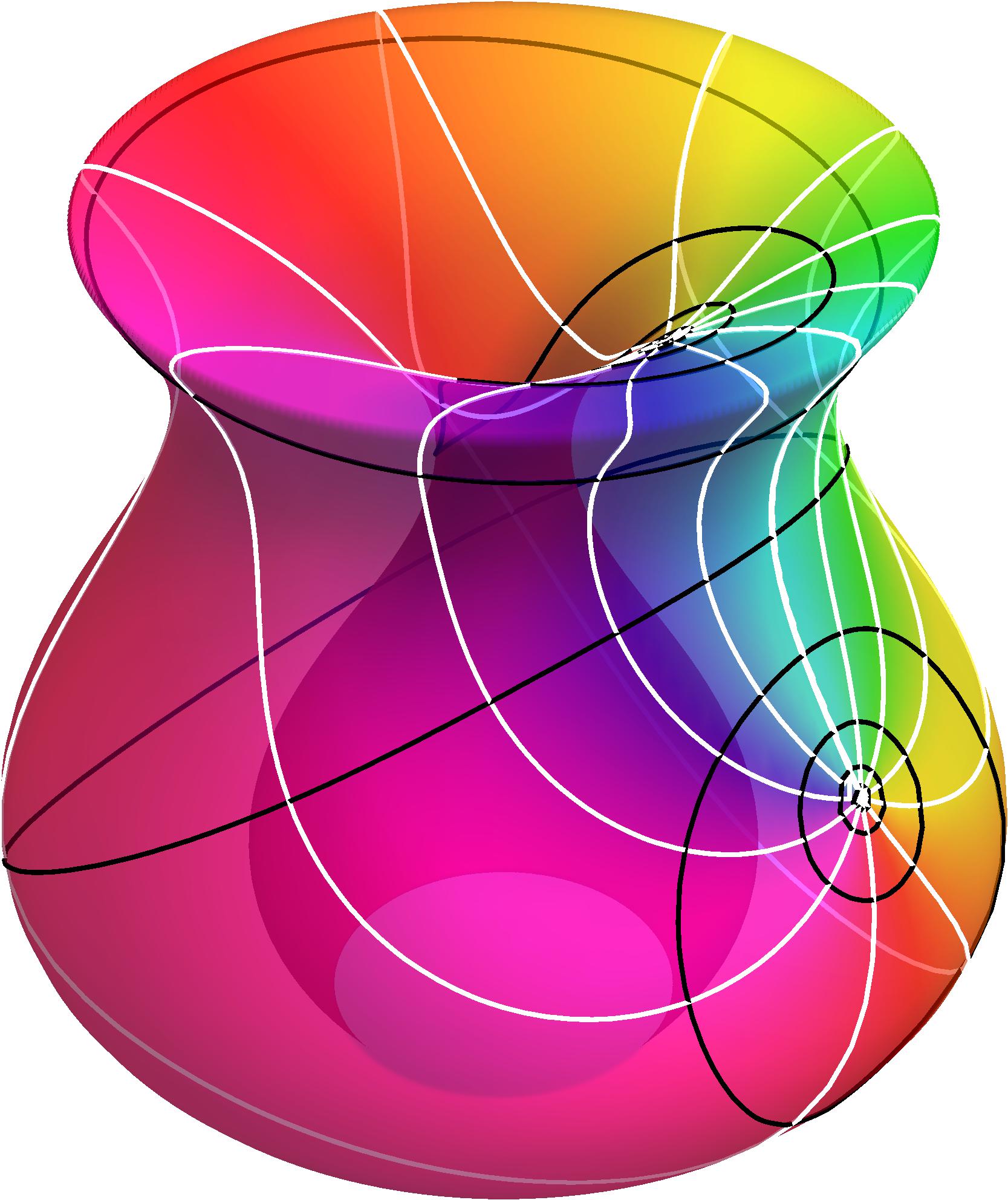}
		\caption{\textit{Left panel:} a closed line in the $x,z$ plane described by ${\bm f}(\eta) = d \{4 \cos \eta -\cos 2 \eta,10\sin\eta - 2 \sin 3 \eta \}$, where $d$ is a length that sets the physical size of the surfaces. \textit{Center panel:} toroid generated by rotating the curve $\bm f$ around a central axis with $a=6\gamma$, see Eq.~\eqref{Revolution_Parametrization}. \textit{Right panel:} The phase pattern induced by a vortex dipole on the surface, via the mapping Eq.~\eqref{IsothermalMap_RS}, with $\gamma=d$, and the flow potential Eq.~\eqref{Potential_Monopole_RS}. \label{fig:RevolutionSurface}
		}
	\end{centering}
\end{figure*}

As before, for a description of a flow field via a complex potential, we must
find an isothermal coordinate set. Let us define $\{u,v\}$ such that $\eta(u,v)=\eta(v)$ only and $\phi(u,v)=\phi(u)$ only. Inserting this ansatz into Eq.~\eqref{LineElement_RS}, we find
\begin{align}
ds^2 = \lambda_\phi^2 \left(\frac{d\phi}{d u}\right)^2 \left(du^2 + \frac{\lambda_\eta^2 \left(\frac{d \eta}{dv}\right)^2}{\lambda_\phi^2 \left(\frac{d \phi}{du}\right)^2} dv^2 \right). \label{LineElement_RS_Map}
\end{align}
For the coordinate set $\{u,v\}$ to be isothermal, the factor in the second term in Eq.~\eqref{LineElement_RS_Map} must equal unity:
\begin{align}
\lambda_\phi(\eta) \frac{d\phi}{d u} =& \lambda_\eta(\eta) \frac{d\eta}{dv} \label{RS_Isothermal_Condition}.
\end{align}
Since $\lambda_\phi,\lambda_\eta$ depend on $\eta$ only, we assume $\phi= u/\gamma$, with $\gamma$ some arbitrary length scale. Then Eq.~\eqref{RS_Isothermal_Condition} can be solved for $v(\eta)$ by separation of variables 
\begin{align}  
\frac{\lambda_\eta(\eta)}{\lambda_\phi(\eta)}  d\eta =& \frac{1}{\gamma} \ dv, \quad\hbox{which gives} \nonumber \\
 v(\eta) =&\, \gamma \int_0^\eta \frac{\sqrt{f_1'(\eta')^2+f_2'(\eta')^2}}{a+f_1(\eta')} d\eta' \label{IsothermalMap_RS}.
\end{align}
The resulting line element is
\begin{align}
ds^2 = \lambda_f(v)^2 \left(du^2+dv^2\right)\label{LineElement_RS_uv},\ \lambda_f(v) = \frac{a+f_1[\eta(v)]}{\gamma}
\end{align}
where $\eta(v)$ is the inversion of Eq.~\eqref{IsothermalMap_RS}. 
The integrand in Eq.~\eqref{IsothermalMap_RS} is always positive, so that $v(\eta)$ is bijective and the inverse $\eta(v)$ well defined. In practice, as long as the length $\int_{-\pi}^\pi |{\bm f(\eta)}'|d\eta$ is finite, both $v(\eta)$ and $\eta(v)$ can easily be evaluated numerically. 

For a standard torus with $\bm f(\eta) = b \{\cos \eta,\sin \eta \}$

 and $\gamma = c = \sqrt{a^2-b^2}$, one recovers Kirchhoff's  transformation Eq.~\eqref{KirchhoffTransformation}.

\subsection{Flow potential}
In the $\{u,v\}$ coordinate set, the toroidal surface of revolution 
is a doubly periodic cell with dimensions $2 \pi \gamma \times \Delta v$, with $\Delta v = v(\pi)-v(-\pi)$. Considering a superfluid confined to the surface described by Eq.~\eqref{Revolution_Parametrization}, the problem of finding the flow potential of vortex configurations in the $\{u,v\}$ plane is identical to that of the  torus: 
\begin{align}
F_{f}(w) =& \sum_n q_n F_{f}(w,w_n) \nonumber \\
F_f(w,w_n)=& \log \jt_1\left(\frac{w-w_n}{2 \gamma},p_f\right)-\frac{\Re\{w_{n}\}}{\gamma \Delta v} w \label{Potential_Monopole_RS},
\end{align}
with $p_f=\exp(-\Delta v/2\gamma)$.

All results obtained in the previous section still apply, 
 with the generalized scaling factor $\lambda_f$   in Eq.~\eqref{LineElement_RS_uv} replacing the scaling factor $\lambda(v)$. 
 Figure~\ref{fig:RevolutionSurface} gives an  example of a toroidal surface of revolution and the corresponding phase pattern of a vortex dipole on its surface.

\section{Outlook}

The superfluid order parameter must be single valued, which leads to the quantization of vortices in two-dimensional planar superfluids. On a cylinder, torus and similar closed surfaces, the same mechanism quantizes flow along any closed path (possibly encircling holes of the surface), significantly altering the flow field of vortices compared to both a classical fluid and a thin cylindrical superfluid film. 
Correspondingly, these restrictions greatly affect  the vortex-vortex interaction energy and the dynamics of the vortices themselves. A spatially-varying metric generates additional contributions to the vortex energy and dynamics.

The inter-vortex energy plays a large role in the transition to the superfluid phase at low temperatures, which is of  Berezinskii-Kosterlitz-Thouless (BKT)  type in two dimensions~\cite{Minnhagen1987}.  
The thermodynamics of vortex excitations in non-planar geometries merits further study \cite{Machta1989}.
Since the quantization of point vortices requires phase coherence on a length scale much shorter than that required for the quantization of flow around the torus circumferences, the vortices on a torus might prove interesting for studying finite-temperature effects and the predicted coherence length in BKT-type superfluids.

Experimentally, the realization of a toroidal superfluid seems difficult. In the context of cold gases, Bose-Einstein Condensates (BECs) have been created in a multitude of trap shapes with unprecedented control over the geometry \cite{Chakraborty2016,Gaunt2013,Lundblad2019_arxiv}. Toroidal shell traps can be engineered for the gas, but the effect of gravity may lead to inhomogeneities in the thickness of the fluid. A possible approach 
would be to do 
the experiment under microgravity. 
Such conditions  may be obtained, for  example, in a free-falling ``Einstein elevator" \cite{Condon2019}, or aboard the International Space Station, where a cold-atom experiment was recently deployed.
Current  proposals  indeed envision the creation of a thin ellipsoidal shell of BEC in space \cite{Lundblad2019_arxiv}. Following our discussion in Sec.~\ref{sec:GenralizedRS}, we expect the quantum effect on vortex dynamics to be robust with regards to the specific shape of the trap geometry, as long as it is toroidal.

Various techniques for optical imprinting of vortices have been proposed \cite{Dobrek1999,Nandi2004}. In a toroidal shell BEC, these techniques should still apply.  
Since an imprinting beam will always pierce the BEC twice, it will  create a vortex anti-vortex pair, consistent with the condition of zero net vorticity. Otherwise, vortices may be formed by rapidly lowering the temperature of the fluid below the critical temperature for superfluidity \cite{Freilich2010}.

\vspace{1cm}
\begin{acknowledgments}
N.G. is supported by a ``la Caixa-Severo Ochoa'' PhD fellowship. As part of the Maciej Lewenstein group, N.G. also acknowledges the Spanish Ministry MINECO (National Plan
15 Grant: FISICATEAMO No. FIS2016-79508-P, SEVERO OCHOA No. SEV-2015-0522, FPI), European Social Fund, Fundació Cellex, Generalitat de Catalunya (AGAUR Grant No. 2017 SGR 1341 and CERCA/Program), ERC AdG OSYRIS and NOQIA, EU FETPRO QUIC, and the National Science Centre, Poland-Symfonia Grant No. 2016/20/W/ST4/00314. 
P.M. is funded by the ``Ram\'on y Cajal" program, and further acknowledges support from the Spanish MINECO (FIS2017-84114-C2-1-P).

\end{acknowledgments}

\bibliography{VortexDipole}
\appendix

\section{Local flow field at a Vortex core}\label{app:LocalFlowApproximation}

In this appendix, we first  study vortex dynamics for  general isothermal  coordinates $\bm s = \{u,v\}$ with metric parameter $\lambda (\bm s)$.   A  torus then represents a simple example with $\lambda$ depending only on $v$.

\subsection{Isothermal coordinates in two dimensions}
Consider  a two-dimensional  surface with an
 isothermal parametrization $\bm{s}=\{u,v\}$  [compare Eq.~(\ref{LineElement_uv})] and metric parameter $\lambda(\bm s)$. To make contact with Sec.~\ref{sec:Dynamics}, it is helpful to focus on the flow near a single vortex core at position $\bm{s}_n$ with charge $q_n$ [included in the definition of the total stream function $\chi(\bm s)$].  

It is convenient to start from the corresponding full complex potential $F(w)$, which is logarithmically singular near the complex coordinate $w_n = u_n + i v_n$.  Separating out this singular part gives the residual regular part 
\begin{equation}\label{reg}
F_n^{\rm reg}(w) = F(w) - q_n \log(w-w_n)
\end{equation}
near $w_n$.

For a general system of well-separated vortices, we now assume that the 
  complex potential $F(w)$ is 
 a sum over individual vortices, as in Eq.~(\ref{FlowPotential_Multipole}).  In the vicinity of the $n$th vortex  $w\approx w_n$, all terms for $m\neq n$ remain finite. 

 Thus the regular contribution to the complex potential  near $w_n$ has the intuitive form

\begin{align}\label{Freg}
F_n^{\rm reg}(w) =& \sum_{m\neq n}\,q_m F(w,w_m)  \\ \nn
   & +\,q_n\,\left[F(w,w_n) - \log(w-w_n)\right].
\end{align}
By construction,  $F_n^{\rm reg}(w)$ remains finite as $ w \to w_n$, with a well-defined value $F_n^{\rm reg}(w_n)$ that includes the contribution from all the other vortices.

Similarly, $\chi(\bm s) = \Re\{F(w)\}$ in the vicinity of $\bm s_n$ separates into  a singular part $q_n \log|\bm s - \bm s_n|$ and 
a regular part  
\begin{equation}\label{chireg}
\chi_n^{\rm reg}(\bm s) 
= \sum_{m\neq n} q_m \chi_m(\bm s) + q_n\left[\chi_n(\bm s) -\log|\bm s - \bm s_n|\right]
\end{equation} 
where $\chi_m(\bm s) = \Re\{F(w,w_m)\}$.
 As for the complex potential, Eq.~(\ref{chireg})  includes the contribution from all the other vortices.    For a flow field described by
such a stream function $\chi(\bm s)$,
 the associated hydrodynamic velocity field is
\begin{align}
\mathbf{w}(\bm s) = \frac{\hbar}{M}\frac{1}{\lambda(\bs)}\hat{\bn} \times \tilde{\bm \nabla} \chi(\bs),
\end{align}
where $\tilde{\bm \nabla} = \bm u\, \partial_u + \bm v \,\partial_v$, as in Eq.~(\ref{Gradient_NaturalBasis}).

Near the vortex core at $\bm s_n$, we   write  $\hat{\bn}\times \tilde{\bm \nabla}\chi(\bs) =  q_n \hat{\bn} \times (\bs-\bs_n)/|\bs-\bs_n|^2  + \hat{\bn} \times \tilde{\bm \nabla}\chi_n^{\rm reg}(\bs) $.  
An   expansion of $\mathbf{w}(\bm s)$ in powers of $\bs-\bs_n$ gives
\begin{align}
\mathbf{w}(\bm s) =& \frac{\hbar}{M}\left[\frac{1}{\lambda(\bs_n)}+ \tilde{\bm \nabla}\left(\frac{1}{\lambda(\bs)}\right)_{\bs\rightarrow\bs_n} \bm \cdot (\bs-\bs_n)\right]\times \nn \\
&\times \left[q_n\frac{\hat{\bn} \times (\bs-\bs_n)}{|\bs-\bs_n|^2} +\hat{\bn} \times \tilde{\bm \nabla}\chi_n^{\rm reg}(\bs_n)\right]+\mO(\bs-\bs_n). \label{Velocity_Local_1}
\end{align}

Introducing the normalized vector $\hat{\bs} = (\bs-\bs_n)/|\bs-\bs_n|$,
we have the leading terms
\begin{align}
\mathbf{w}(\bm s) =& \frac{\hbar}{M} \frac{1}{\lambda(\bs_n)}\hat{\bm n} \times \left[q_n\,\frac{\hat{\bm s}}{|\bs-\bs_n|}+\tilde{\bm \nabla}\chi_n^{\rm reg}(\bs_n)\right. \nn \\
 &\left. - q_n \frac{\hat{\bm s}}{\lambda(\bs_n)}\left(\tilde{\bm \nabla}\lambda(\bs_n)\cdot \hat{\bm s} \right)   \right] + \mO(\bs-\bs_n). \label{Velocity_Local_Final}
\end{align}
Since the flow field in Eq.~\eqref{Velocity_Local_Final} diverges for $\bs \rightarrow \bs_n$, we cannot directly identify 
$\mathbf{w}(\bs_n)$ as the velocity $\mathbf{w}_n$ of the core. Instead, we define the velocity of the vortex core as the angular average  on a  circle of  small radius $|\bm s-\bm s_n|$ around the core $\bs_n$, with $\langle \cdots \rangle = (2\pi)^{-1}\int _0^{2\pi}\,d\theta\, \cdots$.

This angular average can be applied term by term to Eq.~\eqref{Velocity_Local_Final}. The first term vanishes because $\langle \hat{ s}_k\rangle = 0$, 
  the second term is simply a constant, and the identity $\langle \hat s_k\hat s_l \rangle = \frac{1}{2}\delta _{kl}$ simplifies the last term.
 \  In this way we find the  final result for the translational velocity
\begin{align}
\mathbf{w}_n=& \frac{\hbar}{M} \frac{1}{\lambda(\bs_n)}\hat{\bm n} \times \left[ \tilde{\bm \nabla} \chi_n^{\rm reg}(\bs) -\frac{q_n}{2 \lambda(\bs_N)} \tilde{\bm \nabla} \lambda(\bs)\right]_{\bm s \to \bm s_n}
 \nn \\
=& \frac{\hbar}{M} \frac{1}{\lambda(\bs_n)} \hat{\bn} \times \left[ \tilde{\bm \nabla} \left(\chi_n^{\rm reg}(\bs)-\frac{q_n}{2}\log\lambda(\bs)\right)\right]_{\bm s\to \bm s_n}.
\label{Velocity_Curved_Final}
\end{align} 

In connection with the study of vortex dynamics in Sec.~\ref{sec:Dynamics},  the last equation has the equivalent form 
\begin{align}\label{velocity-curved}
\mathbf{w}_n & =  \frac{\hbar}{M} \frac{1}{\lambda(\bs_n)} \nn \\ 
& \times \left[ \left(\bm v\frac{\partial }{\partial u} -\bm u\frac{\partial}{\partial v}\right) 
  \left(\chi_n^{\rm reg}(\bs)-\frac{q_n}{2}\log\lambda(\bs)\right)\right]_{\bm s\to \bm s_n}.
\end{align}
Here, the Hamiltonian structure identifies  $u_n$ and $v_n$ as the canonical variables.
 
 \subsection{Application to a torus}
 
 For a torus, the second part of $F_n^{\rm reg}(w)$ in Eq.~(\ref{Freg}) is $q_n \log\left\{\vartheta_1[(w-w_n)/2c,p]/(w-w_n)\right\}-q_nu_n w/(2\pi bc)$. Here, the first term is even in its first argument $w-w_n$ and hence does not contribute to the vortex velocity.   As a result, the regular part of the complex potential  can be taken as 
  \begin{align}\label{Fregular}
\bar{F}_n^{\rm reg}(w) =& \sum_{m\neq n} q_m  \log\vartheta_1\left(\frac{w-w_m}{2c},p\right) 
-\sum_m \frac{q_mu_m w}{2\pi bc}. 
\end{align}
Correspondingly, the regular part of the stream function can be taken as $\bar{\chi}_n^{\rm reg}(\bm s) = \Re\{ \bar{F}_n^{\rm reg}(w)\}$.

As for a plane, the two real vector components  in Eq.~(\ref{velocity-curved}) combine to give the complex expression in Eq.~(\ref{Velocity_Vortex_Torus}),
\begin{align}
\Omega_n =&\,{\bf w}_{n,v} + i {\bf w}_{n,u} \nn \\
=& \frac{\hbar}{M}\frac{1}{ \lambda(v_n)}\bigg( \sum_{m\neq n}q_m f(w_n,w_m) \nn \\
 &  + i \frac{q_n}{2} \frac{\lambda'(v_n)}{\lambda(v_n)} - \ q_n \frac{u_n}{2 \pi b c} \bigg)
\end{align} 
with $f(w_n,w_m)$ defined by Eq.~\eqref{Velocity_Monopole}. 

\end{document}